\documentclass[a4paper,fleqn,usenatbib]{mnras}
\usepackage{graphics,graphicx,rotating}
\usepackage{natbib}
\usepackage{xspace}
\usepackage{amssymb}
\usepackage{amsmath}
\usepackage{epstopdf}
\usepackage{subfigure}
\usepackage{pdflscape}
\usepackage{rotating}
\usepackage{xcolor}

\begin{document}
\title[Large depleted core in IC~1101.] {A remarkably large depleted core in the Abell 2029 BCG IC 1101 }
\author[Dullo, Graham \& Knapen] {Bililign T.\
  Dullo$^{1,2,3}$\thanks{bdullo@ucm.es},  Alister W.\ Graham$^{4}$ and
  Johan H.\ Knapen$^{2,3}$\\
$^{1}$Departamento de   Astrof\'isica y Ciencias de
    la Atm\'osfera, Universidad Complutense de Madrid, E-28040 Madrid,
    Spain\\
$^{2}$Instituto de Astrof\'isica de Canarias, V\'ia L\'actea S/N, E-38205 La Laguna, Spain\\
$^{3}$Departamento de Astrof\'isica, Universidad de La Laguna, E-38206 La Laguna, Spain\\
$^{4}$Centre for Astrophysics and Supercomputing, Swinburne University of
  Technology, Hawthorn, VIC 3122, Australia}
 \maketitle
\label{firstpage}
\begin{abstract}
 
  We report the discovery of an extremely large ($R_{\rm b} \sim$
  $2\farcs77 \approx$ 4.2 kpc) core in the brightest cluster galaxy,
  IC 1101, of the rich galaxy cluster Abell 2029. Luminous
  core-S\'ersic galaxies contain depleted cores---with sizes
  ($R_{\rm b}$) typically 20 -- 500 pc---that are thought to be formed
  by coalescing black hole binaries. We fit a (double nucleus) +
  (spheroid) + (intermediate-scale component) + (stellar halo) model
  to the {\it Hubble Space Telescope} surface brightness profile of
  IC~1101, finding the largest core size measured in any galaxy to
  date. This core is an order of magnitude larger than those typically
  measured for core-S\'ersic galaxies. We find that the spheroid's
  $V$-band absolute magnitude ($M_{V}$) of $-23.8$ mag ($\sim $25\% of
  the total galaxy light, i.e., including the stellar halo) is faint
  for the large $R_{\rm b}$, such that the observed core is 1.02 dex
  $\approx$ 3.4$\sigma_{s}$ (rms scatter) larger than that estimated
  from the $R_{\rm b} -M_{V}$ relation. The suspected scouring process
  has produced a large stellar mass deficit ($M_{\rm def}$) $\sim$
  $4.9 \times 10^{11} M_{\sun}$, i.e., a luminosity deficit $\approx$
  28\% of the spheroid's luminosity prior to the depletion. Using IC
  1101's black hole mass ($M_{\rm BH}$) estimated from the
  $M_{\rm BH}-\sigma$, $M_{\rm BH}-L$ and $M_{\rm BH}-M_{*}$
  relations, we measure an excessive and unrealistically high number
  of ``dry" major mergers for IC~1101 (i.e., $\mathcal{N} \ga 76$) as
  traced by the large $M_{\rm def}/M_{\rm BH}$ ratios of $38-101$. The
  large core, high mass deficit and oversized $M_{\rm def}/M_{\rm BH}$
  ratio of IC~1101 suggest that the depleted core was scoured by
  overmassive SMBH binaries with a final coalesced mass
  $M_{\rm BH} \sim (4 -10) \times 10^{10} M_{\sun}$, i.e., $\sim$ (1.7
  -- 3.2$)\times\sigma_{s}$ larger than the black hole masses
  estimated using the spheroid's $\sigma$, $L$ and $M_{*}$. The large
  core might be partly due to oscillatory core passages by a
  gravitational radiation-recoiled black hole.

\end{abstract}


\section{Introduction}

Many luminous early-type galaxies brighter than
$M_{B} \sim- 20.5 \pm 0.5$ mag contain partially-depleted cores. Given
their high luminosity, the brightest cluster galaxies (BCGs) are
therefore interesting targets for investigating the actions of
supermassive black holes (SMBHs). Ground-based observations first
revealed these depleted cores as a flattening in the inner stellar
light distributions (e.g., \citealt{1966ApJ...143.1002K};
\citealt{1978ApJ...222....1K}; \citealt{1994MNRAS.268L..11Y};
\citealt{1982MNRAS.200..361B}). High-resolution imaging with the {\it
  Hubble Space Telescope (\it HST)} subsequently allowed us to
characterise cores robustly, including those that are small in angular
size and unresolved from the ground (e.g.,
\citealt{1993AJ....106.1371C}; \citealt{1994ESOC...49..147K};
\citealt{1994AJ....108.1567J}; \citealt{1994AJ....108.1598F};
\citealt{1994AJ....108..102G}; \citealt{1994AJ....108.1579V};
\citealt{1995AJ....109.1988F}; \citealt{1995AJ....110.2622L};
\citealt{1996AJ....111.1889B}; \citealt{1996AJ....112..105G};
\citealt{1997AJ....114.1771F}; \citealt{2001AJ....122..653R};
\citealt{2001AJ....121.2431R}; \citealt{2003AJ....125..478L};
\citealt{2006ApJS..164..334F}; \citealt{2009ApJ...691L.142K};
\citealt{2012ApJ...755..163D, 2013ApJ...768...36D,
  2014MNRAS.444.2700D}; \citealt{2013AJ....146..160R};
\citealt{2013ARA&A..51..511K}).

Here, we report the discovery of an extremely large ($R_{\rm b} \sim$
$2\farcs77 \pm 0.07 \approx$ 4.2 $\pm 0.1$ kpc) core, in fact, the
largest to date, in the Abell 2029 cluster's BCG IC 1101 by fitting
the core-S\'ersic model to the {\it HST} light profile of the
spheroid. The core-S\'ersic model describes the light profile of
luminous ``core-S\'ersic galaxies'' which have an inner light profile
with a shallow negative logarithmic slope\footnote{\citet[see also
  \citealt{1997AJ....114.1771F};
  \citealt{2001AJ....121.2431R}]{1995AJ....110.2622L} defined the
  inner logarithmic slope of the galaxy inner light profiles as
  $-\gamma$.}  ($\gamma \la 0.3$) that deviates downward relative to
the inward extrapolation of the outer \citet{1968adga.book.....S}
$R^{1/n}$ profile (e.g., \citealt{2003AJ....125.2951G};
\citealt{2004AJ....127.1917T}; \citealt{2006ApJS..164..334F};
\citealt{2012ApJ...755..163D, 2013ApJ...768...36D,
  2014MNRAS.444.2700D}).

Recent works advocated measuring depleted cores of galaxies using the
``cusp radius ($r_{\gamma=0.5}$)'' which was introduced by
\citet{1997ApJ...481..710C} and equals the radius where the negative
logarithmic slope of the Nuker model profile equals 0.5 (e.g.,
\citealt{2007ApJ...662..808L}; \citealt{2012ApJ...756..159P};
\citealt{2014ApJ...795L..31L}). While the cusp radius agrees better
with the core-S\'ersic radius than the Nuker break radius
(\citealt{2007ApJ...662..808L}), \citet{2012ApJ...755..163D,
  2013ApJ...768...36D} warned that the light profile of a coreless
galaxy can have a radius where $\gamma =0.5$.

Cores measured using the core-S\'ersic model have typical sizes
$R_{\rm b} \sim 20 - 500$ pc (e.g.,\citealt{2004AJ....127.1917T};
\citealt{2006ApJS..164..334F}; \citealt{2011MNRAS.415.2158R};
\citealt{2012ApJ...755..163D, 2013ApJ...768...36D,
  2014MNRAS.444.2700D}; \citealt{2013AJ....146..160R}), compared to
$R_{\rm b}\sim$ 4.2 kpc for IC~1101. \citet{2014ApJ...795L..31L}
reported an extremely large ($r_{\gamma=0.5}$ $\approx$ 4.57 kpc) core
for the Abell 85 cluster's BCG Holm 15A. However, this result has
since been questioned (\citealt{2015ApJ...807..136B};
\citealt{2016ApJ...819...50M}). \citet{2012ApJ...756..159P} fit the
Nuker model (\citealt{1995AJ....110.2622L}) to the major-axis light
profile of the Abell 2261 BCG and found the hitherto largest core
($r_{\gamma=0.5} \sim $ 3.2 kpc), which was recently confirmed by
\citet{2016ApJ...829...81B} who fit a 2D core-S\'ersic model to the
galaxy's image and measured $R_{\rm b} \sim 3.6$ kpc. The Abell 2261
BCG core is over a factor of two larger than the largest core in the
\citet{2007ApJ...662..808L} sample of galaxies ($r_{\gamma=0.5} \sim$
1.5 kpc for NGC 6166, see also \citealt{2003AJ....125..478L}).

Theory predicts that central stellar mass deficits are created when a
coalescing SMBH binary ejects stars with a collective mass that is on
the order of the SMBH mass from the centre of a newly formed ``dry''
(gas-poor) galaxy merger remnant via a three-body gravitational
scattering process (e.g., \citealt{1980Natur.287..307B};
\citealt{1991Natur.354..212E}; \citealt{2001ApJ...563...34M};
\citealt{2006ApJ...648..976M}). Large depleted cores of galaxies may
be generated by ultramassive ($\ga 10^{10}$ $M_{\sun}$) BH binaries
and/or the cumulative actions of SMBH binaries that are created during
multiple successive dry mergers (e.g., \citealt{2001ApJ...563...34M};
\citealt{2006ApJ...648..976M}). The energy liberated from the
orbitally decaying SMBH binary typically ejects stars that are on
radial orbits, leaving a relative excess of tangential orbits in the
galaxy cores (e.g., \citealt{1997NewA....2..533Q};
\citealt{2001ApJ...563...34M}; \citealt{2003ApJ...583...92G};
\citealt{2014ApJ...782...39T,
  2016Natur.532..340T}). \citet{2015ApJ...798...55D} showed that these
core regions tend to be round.

Additional core depletion mechanisms can enlarge the stellar mass
deficit and the depleted core of galaxies. For example, the SMBH that
is produced from the coalesce of the inspiralling SMBH binary may
recoil to conserve the linear momentum that is carried away in the
other direction by the anisotropic emission of gravitational wave
radiation (\citealt{1973ApJ...183..657B};
\citealt{1983MNRAS.203.1049F}; \citealt{1989ComAp..14..165R}). Most
gravitational wave-recoiled SMBHs have kick velocities lower than
their host galaxies' escape velocities (e.g.,
\citealt{2008ApJ...678..780G}). Therefore, the recoiled SMBH ejects
yet more stars as it repetitively oscillates in the core region of its
host, resulting in a larger depleted core/mass deficit
(\citealt{2004ApJ...607L...9M}; \citealt{2004ApJ...613L..37B};
\citealt{2008ApJ...678..780G}). If the coalesced SMBH does escape the
galaxy, then the core expands as it is no longer so tightly bound.
Furthermore, \citet{2012MNRAS.422.1306K} proposed that the scouring
action of multiple SMBHs generates larger stellar mass deficits than a
single SMBH binary due to the SMBH-SMBH encounters. This scenario
assumes that multiple SMBHs are present in galaxies at high redshift
because of high galaxy merger rates.

Using numerical simulations that did not include SMBHs,
\citet{2010ApJ...725.1707G} proposed core generation by a stalled
perturber that is captured and dragged to the centre of a galaxy
(e.g., \citealt{2015ApJ...806..220A}). In this scenario, the energy
transferred from the perturber to the stars at the centre of the
galaxy would produce a core (see also \citealt{2012ApJ...745...83A}  and
\citealt{2016MNRAS.456.2457A}).

 \begin{figure}
\begin {minipage}{85mm}
\vspace*{-.04414643199cm}   
\hspace*{-.5043199cm}   
 \includegraphics[angle=0,scale=0.310035062]{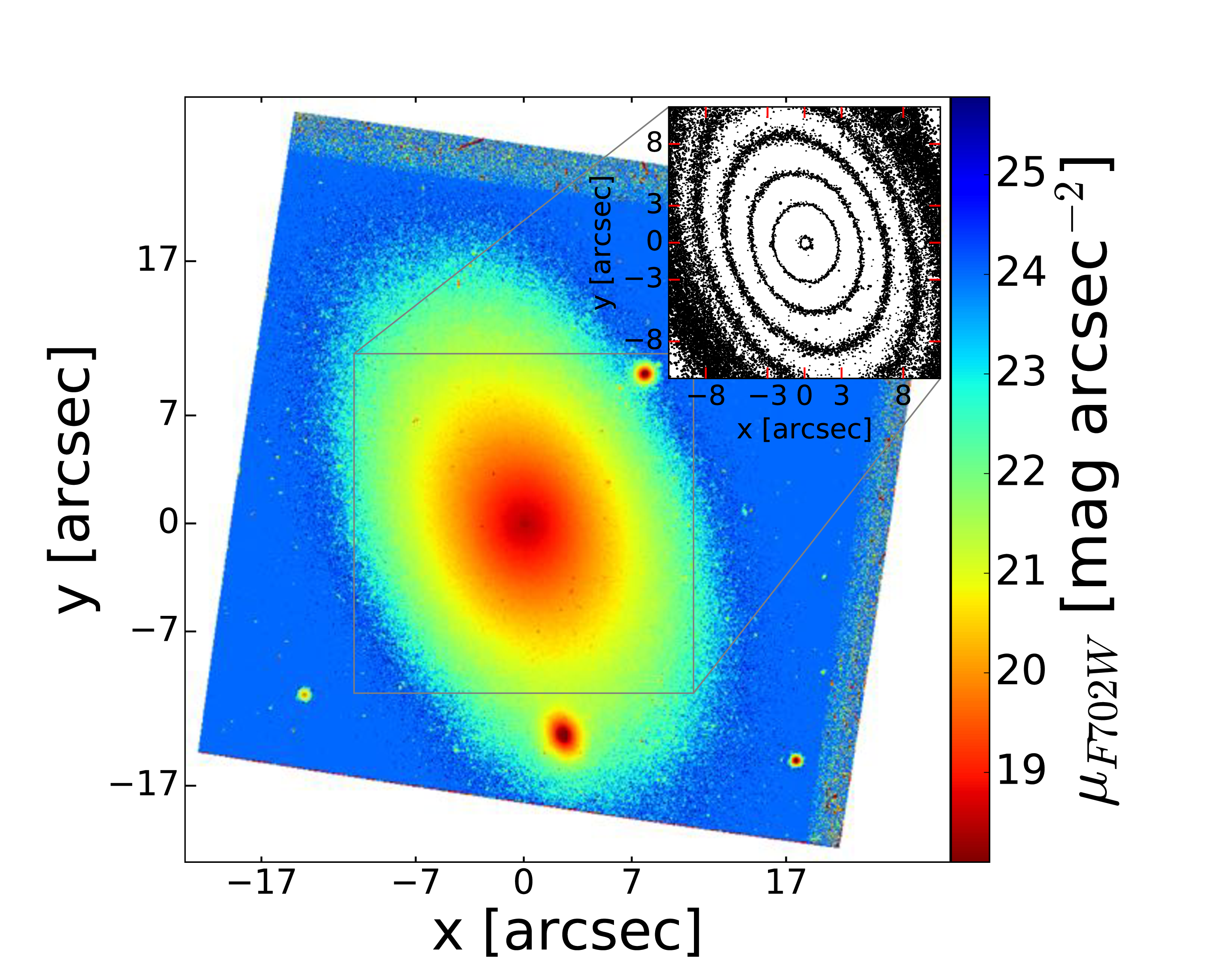} 
 \caption{{\it HST} PC1 $F702W$ image of the A2029-BCG IC 1101,
   showing the large depleted core with break radius surface
   brightness of $\mu_{\rm b,{\it F702W}} \sim$ 19.4 mag
   arcsec$^{-2}$. The top inset shows the contour levels 18.6, 19.3,
   20.1, 20.8, 21.6, 22.5, 23.3 $\times$ mag arcsec$^{-2}$). North is
   up and east is to the left. }
\label{Fig1} 
\end {minipage}
\end{figure}  

\begin{center}

\begin{table} 
\setlength{\tabcolsep}{0.15548in}
\begin {minipage}{84mm}
\caption{IC~1101 basic data}
\label{Tab1}
\begin{tabular}{@{}llcccc@{}}
\hline
\hline
Galaxy&Type& $B_{T}$  &$\sigma $&$z$\\
&&(mag)&(km s$^{-1}$)& \\
(1)&(2)&(3)&(4)&(5)&\\
\multicolumn{1}{c}{} \\              
\hline                           

IC 1101    &  BCG   &15.3&378&0.0799 \\

\hline
\end{tabular} 

Notes.---Col.\ 1: galaxy name. Col.\ 2: morphological type, the galaxy
is classified as an S0 in the RC3 (\citealt{1991rc3..book.....D}).
Cols.\ 3 - 4: total $B$-band magnitude and central velocity dispersion
from HyperLeda (http://leda.univ- lyon1.fr;
\citealt{2003A&A...412...45P}). Col. (5): redshift (see
Section~\ref{Sec1.1}).
\end {minipage}
 \end{table}
\end{center}

\subsection {IC~1101}\label{Sec1.1}

IC~1101 is the BCG of the richness class 4.4 galaxy cluster Abell 2029
(\citealt{1978ApJ...226...55D}) at a redshift of $z$=0.0799
(NED\footnote{We assume a cosmology with $H_{0}$ = 70 km $s^{-1}$
  Mpc$^{-1}$, $\Omega_{\Lambda}$ = 0.7, and $\Omega_{m}$ = 0.3. },
Virgo + GA + Shapley). A2029 is regarded as one of the most relaxed
clusters in the Universe (e.g., \citealt{1979ApJ...231..659D};
\citealt{1996ApJ...458...27B}).

Our adopted redshift for IC~1101 gives a luminosity distance of 363
Mpc and a scale of 1.51  kpc arcsec$^{-1}$.

Section \ref{Sec2} describes the data reduction, the surface
brightness profile and isophotal parameter extraction techniques. In
Section \ref{Sec3}, we discuss the analytic model fit to the light
profiles of IC~1101 together with the analysis of our four-component
decompositions. In Sections \ref{Sec3.2} and \ref{Sec3.3}, we discuss
the colour map and the isophotal properties, respectively, of IC~1101
in the context of our light profile decompositions.  Section
\ref{Sec4.0} discusses past works on IC~1101 and similar BCGs. Section
\ref{Sec4.1} compares the large depleted core of IC~1101 with those of
other core galaxies in the literature. In Section \ref{Sec4.2}, we
discuss the stellar mass deficit of IC~1101.  Section \ref{Sec4.3}
discusses a possible formation scenario for IC~1101 and Section
\ref{ConV} summarizes our main conclusions.

\section{Data}\label{Sec2}
\subsection{{\it HST} imaging} 

High-resolution {\it Hubble Space Telescope (HST)} Wide-Field
Planetary Camera 2 (WFPC2) and Planetary Camera 1 (PC1) images of IC
1101 taken with $F450W$ and $F702W$ filters were retrieved from the
public Hubble Legacy Archive (HLA)\footnote{http://hla.stsci.edu.}.
These images were obtained under a proposal ID 6228 (PI: Trauger).
Fig.~\ref{Fig1} reveals the remarkably flattened core of IC~1101
imaged with PC1 chip of the WFPC2 camera.

\begin{figure*}
  \begin{tabular}{|c|c|c|}
\hspace{-.69001355cm}
  \includegraphics[width=0.47849115785\textwidth,height=0.4246371085\textwidth]{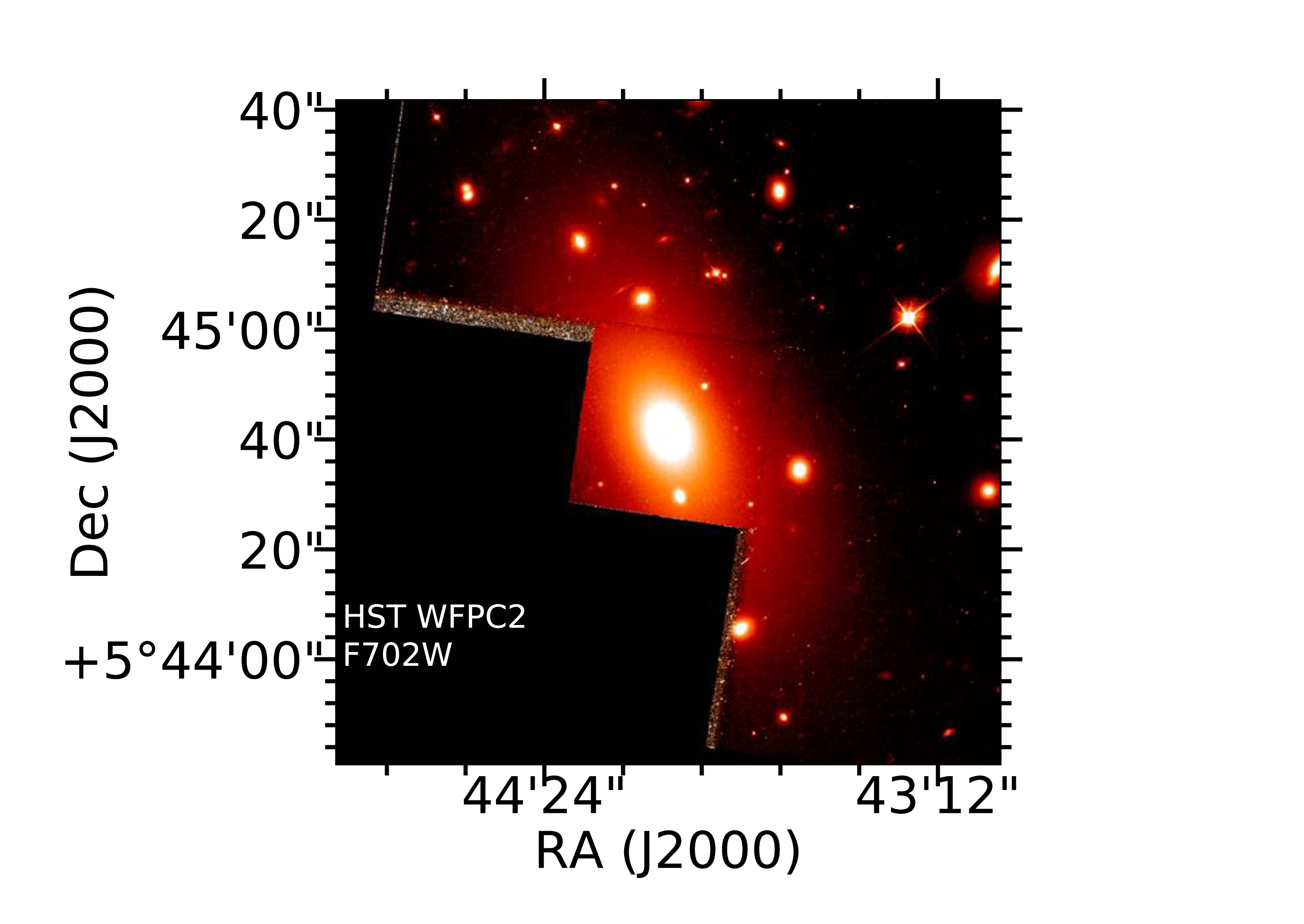}    &
\hspace{-2.349001355cm}
   \includegraphics[width=0.47849115785\textwidth,height=0.4246371085\textwidth]{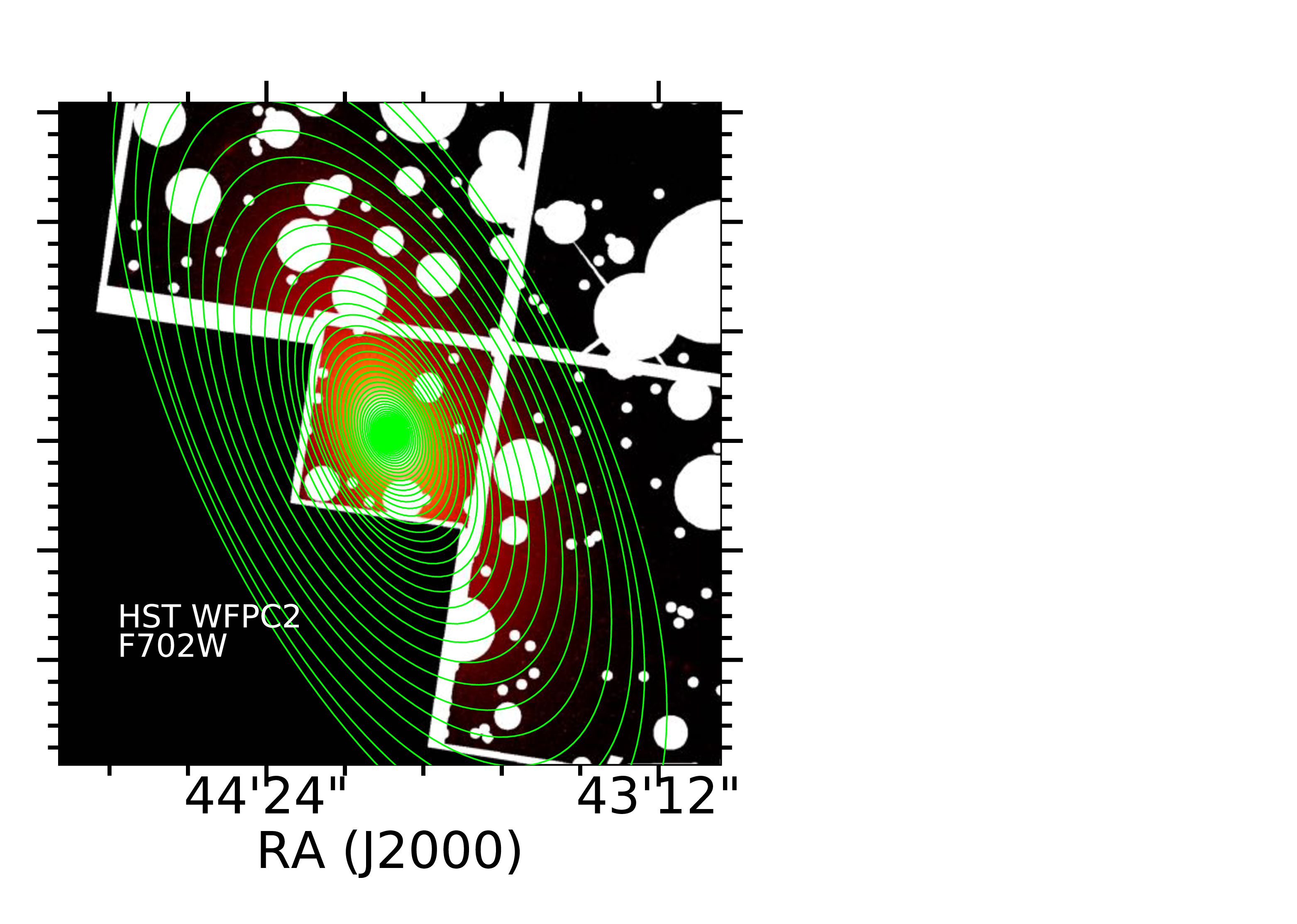} &
\hspace{-3.9989409001355cm}
\vspace{-.5609001355cm}
    \includegraphics[width=0.47849115785\textwidth,height=0.4246371085\textwidth]{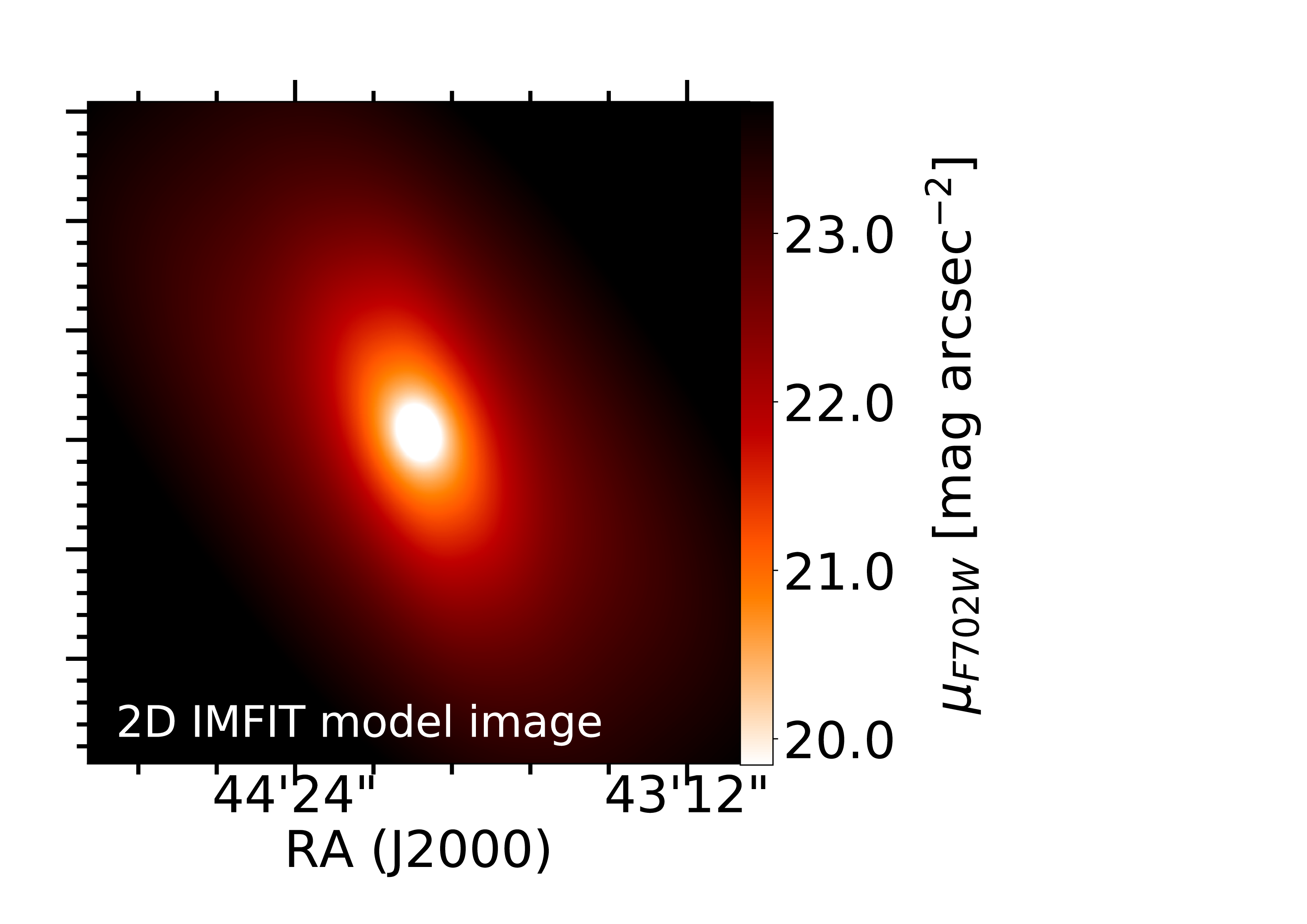}         
  \end{tabular}
\caption{Left-hand panel: {\it HST} WFPC2 F702W image of IC
  1001. Middle panel: masked regions (i.e., white areas) and isophotes
  from IRAF {\sc ellipse} overplotted on the {\it HST}
  image. Right-hand panel: {\sc imfit} model image of IC~1101 (see Section~\ref{Sec3}).}
\label{Fig3} 
\end{figure*}

While the {\it HST} WFPC2 images obtained from the HLA archive have a
160$\arcsec$$\times$160$\arcsec$ L-shaped field-of-view (FOV) at an
image scale of 0$\farcs$1
pixel$^{-1}$, the PC1 images with a plate scale of
0$\farcs$05 pixel$^{-1}$ have a
40$\arcsec$$\times$40$\arcsec$ FOV (Fig.\ref{Fig3}).

\subsection{Surface brightness profiles}\label{Surf_br}

Our 1D data reduction steps, along with the surface brightness profile
extraction procedures follow those in \citet[their Section
2.3]{2014MNRAS.444.2700D}. We used the IRAF\footnote{IRAF is
  distributed by the National Optical Astronomy Observatory, which is
  operated by the Association of Universities for Research in
  Astronomy (AURA) under cooperative agreement with the National
  Science Foundation.} {\sc ellipse} task
(\citealt{1987MNRAS.226..747J}) to extract surface brightness,
ellipticity ($\epsilon$ = 1 - b/a), position angle (P.A.), and
isophote shape parameter (B$_{4}$) profiles from the high-resolution
{\it HST} WFPC2 $F450W$ and $F702W$ images of IC 1101 (see
Fig.~\ref{Fig3}). A mask was created first by running {\sc SExtractor}
(\citealt{1996A&AS..117..393B}) on the galaxy image and then combining
this with a careful manual mask to exclude all objects except the BCG,
and to avoid the gaps between individual CCD detectors and the
partially missing quadrant of the WFPC2 images (Fig.~\ref{Fig3}).  We
ran {\sc ellipse} by holding the isophote centres fixed but allowing
$\epsilon$ and P.A.  to vary. Fig.~\ref{Fig3} (middle) shows the
resulting isophotes. Fig.~\ref{FigRes} (left) shows a residual image
that is produced by subtracting the model image of the ellipse fit
from the science image.

\subsubsection{Sky background}\label{Sky}

We focus on the {\it HST} WFPC2 $F702W$-band light profile of IC~1101
in this work instead of the $F450W$-band light profile to minimize the
effects of young stellar populations and dust contamination, although
this galaxy appears to be dust free.  Fig.~\ref{Fig3} shows that the
outermost corner of the WF3 chip of the WFPC2 is mostly free from
IC~1101's light, suggesting optimum sky-background level determination
by the HLA pipeline. However, because IC~1101 extends beyond the WFPC2
CCDs (Fig.~\ref{Fig3}), poor sky background subtraction can bias the
extracted profile at low surface brightness. Concerned about this, we
extracted light profiles from the mosaic SDSS $r$- and $z$-band images
of IC 1101 (Fig.~\ref{Fig4}), with a
5$\arcmin$$\times5\arcmin$ FOV, using {\sc ellipse}. We found that the
{\it HST} $F702W$ and SDSS
$z$-band profiles agree very well but they differ slightly from the
SDSS
$r$-band profile at large radii (Fig.~\ref{Fig55}). This discrepancy
is due to a poor background subtraction of the SDSS
$r$-band image by the SDSS pipeline, resulting in a non-uniform
background that can be seen in the image as dark and bright stripes
(see \citealt{2011AJ....142...31B}).  Nonetheless, in
Section~\ref{Sec3}, we model a composite ({\it HST}
$F702W$ plus SDSS
$r$-band) profile of the galaxy and find that the discrepancy between
the {\it HST} and SDSS
$r$-band profiles did not affect the conclusions in the paper.

We quote all magnitudes in the VEGA magnitude system.

\section{Decomposing IC~1101} \label{Sec3}
\subsection{1D decomposition}\label{Sec33.1}

We fit a (core-S\'ersic spheroid) plus (S\'ersic intermediate-scale
component) plus (exponential halo) plus (smaller inner Gaussian
component) to the {\it HST} $F702W$ and $F450W$ brightness profiles of
IC~1101 (Figs.~\ref{Fig5}, and \ref{FigMnG}).
 
The \citet{1963BAAA....6...41S} model describes the surface brightness
profiles of low- and intermediate-luminosity ($M_{B}$ $\ga-$ 20.5 mag)
spheroids (see the review by \citealt{2005PASA...22..118G}, and
references therein). This model is defined as
\begin{equation}
I(R) = I_{\rm e} \exp \left[ - b_{n}
\left(\frac{R}{R_{\rm e}}\right)^{1/n} -1\right],
 \label{Eqq2}
\end{equation}
where $ I_{\rm e}$ denotes the intensity at the half light radius
($R_{\rm e}$).
The quantity $b_{n}\approx 2n- 1/3$, for $1\la n\la 10$
is defined as a function of the S\'ersic index $n$ such that $R_{\rm e}$
encloses half of the total luminosity. For $n= 0.5$ and 1, the
S\'ersic model is a Gaussian function and an exponential function,
respectively. 

As noted in the Introduction, luminous (M$_{B} \la -20.5 $ mag)
galaxies posses central stellar deficits such that their inner light
profiles depart downward relative to the inward extrapolation of the
outer S\'ersic profile. The core-S\'ersic model\footnote{Because the
  core-S\'ersic model gives a good description of galaxies with
  depleted cores, we refer to such galaxies as ``core-S\'ersic
  galaxies''.}, a combination of an inner power-law core and an outer
S\'ersic profile with a transition region, describes very well the
underlying light profiles of spheroids with depleted cores. The
core-S\'ersic model is written as
 
\begin{equation}
I(R) =I' \left[1+\left(\frac{R_{\rm b}}{R}\right)^{\alpha}\right]^{\gamma /\alpha}
\exp \left[-b\left(\frac{R^{\alpha}+R^{\alpha}_{\rm b}}{R_{\rm e}^{\alpha}}
\right)^{1/(\alpha n)}\right], 
\label{Eq2}
 \end{equation}
with 
\begin{equation}
I^{\prime} = I_{\rm b}2^{-\gamma /\alpha} \exp 
\left[b (2^{1/\alpha } R_{\rm b}/R_{\rm e})^{1/n}\right], 
 \end{equation}
where $I_{\rm b}$ is intensity at the core break radius $ R_{\rm b}$, $\gamma$ is
the slope of the inner power-law region, and $\alpha$ moderates the
sharpness of the transition between the inner power-law core and the outer
S\'ersic profile. $R_{\rm e}$ and $b$  are defined as in the S\'ersic model.

\begin{figure*}
 \begin{tabular}{|c|c|c|}
\includegraphics[width=0.57478\textwidth]{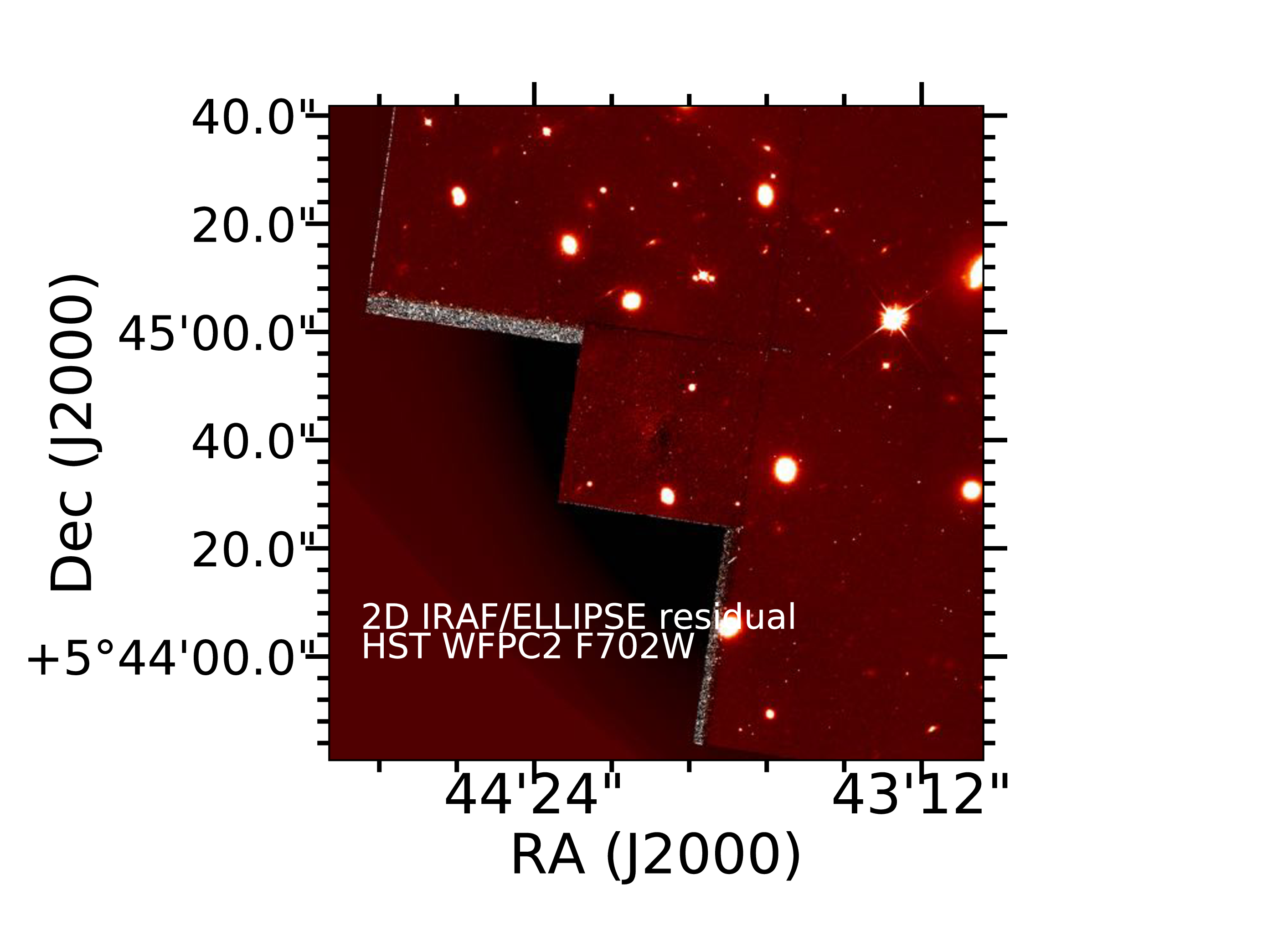}
\hspace{-1.00853409001355cm}
\vspace{-.3809001355cm}
\includegraphics[width=0.57450536\textwidth]{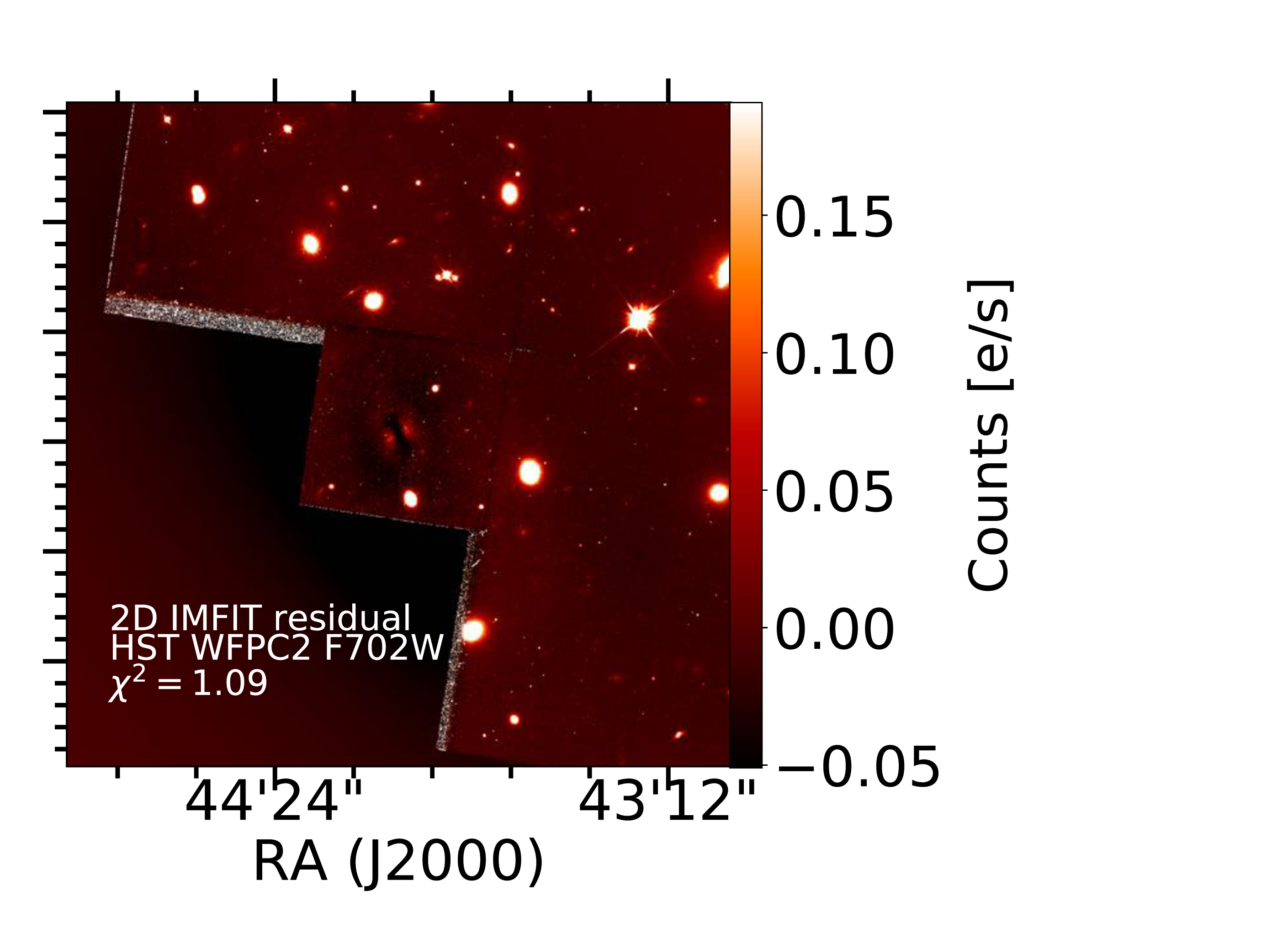}
 \end{tabular}
\caption{Left-hand panel: residual image which is created after
  subtracting a model image (generated by the IRAF {\sc bmodel} task)
  from the original {\it HST} WFPC2 F702W image shows positive and
  negative residuals to the NE and SW of the core, respectively.  This
  suggests that the core is somewhat offset with respect to the outer
  isophotes, consistent with the scenario where the depleted core is
  enlarged through oscillatory core passages by a gravitational
  radiation-recoiled black hole (see \citealt{2012ApJ...756..159P} and
  Section~\ref{Sec4}). The top right and bottom left regions of this
  residual image are outside the IRAF {\sc bmodel} image of ~IC~1101
  (see Fig~\ref{Fig3}, middle panel). Right-hand panel: {\sc imfit}
  residual image. The residual structure inside $R \sim 10\arcsec$ is
  due to IC~1101's ellipticity gradient which is not well represented
  by the {\sc imfit} core-S\'ersic model component with
  $\epsilon~\sim 0.24$ that dominates at $R \la 10\arcsec$ (see the
  text for details). The count rates in the residual images that are
  shown in units of electrons per second and the 2D surface brightness
  $\mu_{F702W}$ (mag arcsec$^{-2}$, Fig~\ref{Fig3}) are related as
  $\mu_{F702W} $= $-$2.5 log (count rates) + 19.547.}
\label{FigRes} 
\end{figure*}

\begin{figure}
\vspace{-.709001355cm}
\hspace{-.679001355cm}
\includegraphics[angle=0,scale=0.23890380295]{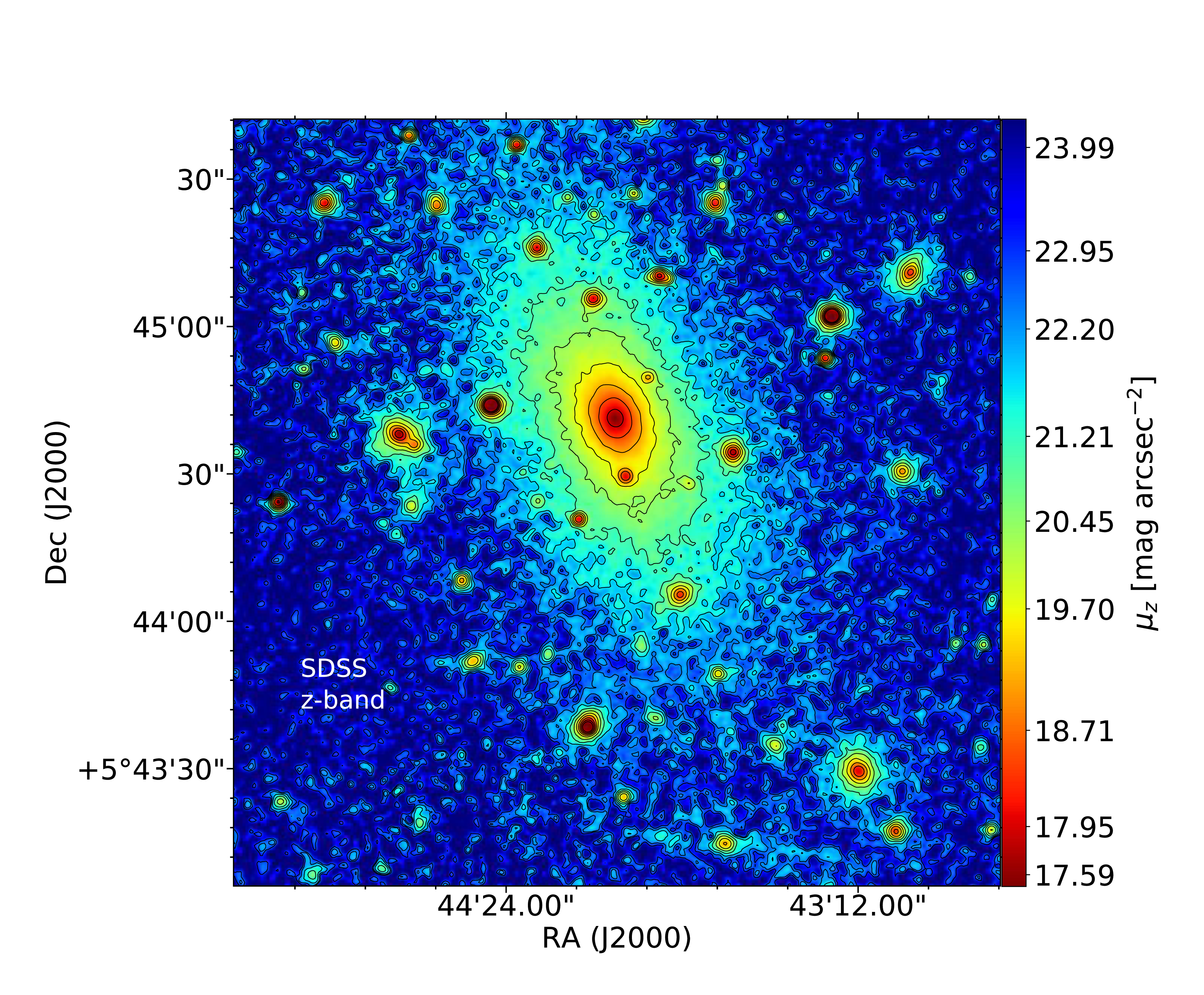}
\caption{SDSS $z$-band image and contour plot of IC~1101 smoothed with a
  Gaussian (FWHM $\sim$1$\farcs$5), highlighting the intermediate-scale and stellar halo
  components.}
 \label{Fig4} 
 \end{figure}

\begin{figure}
\includegraphics[angle=270,scale=.9133]{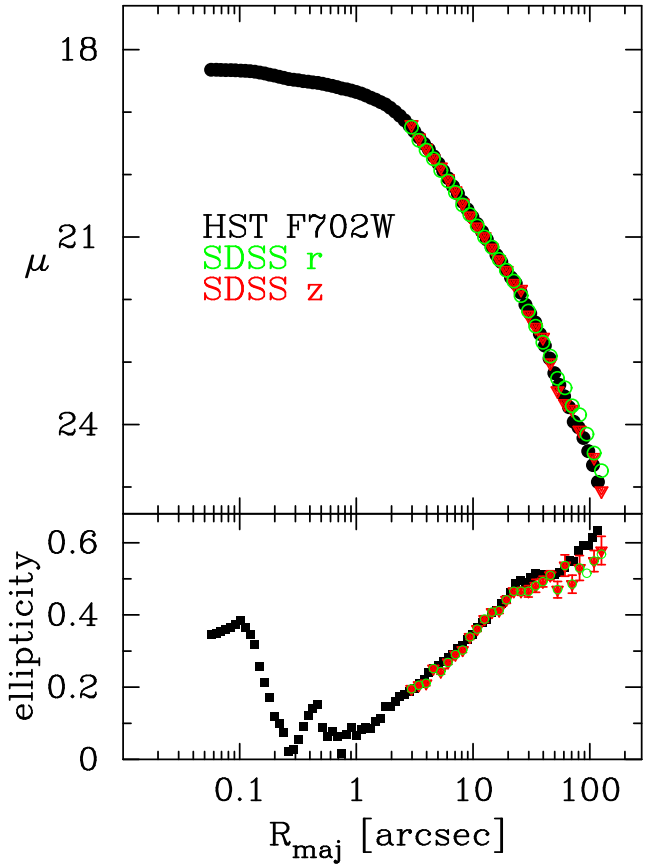}
\caption{Major-axis {\it HST} $F702W$-, and SDSS $z$- and $r$-band surface
  brightness and ellipticity profiles of IC~1101 denoted by black dots, red
  triangles and open green circles, respectively. The SDSS $r$- and
  $z$-band data are zero-pointed to the {\it HST} $F702W$ profile. For
  clarity, we only show the error bars on the SDSS $z$-band ellipticities. }
\label{Fig55} 
 \end{figure}

\begin{figure*}
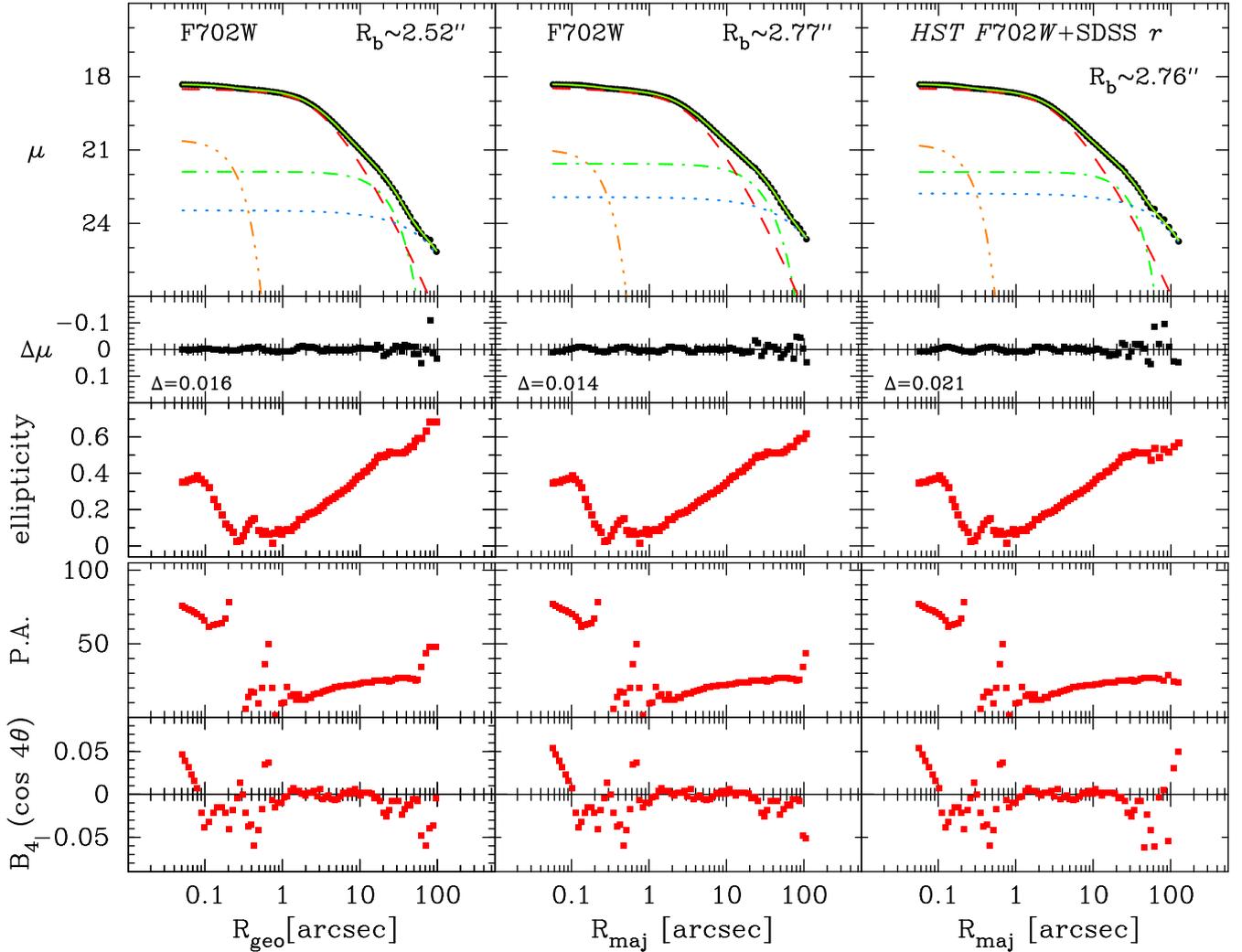

\hspace*{-0.3262527cm}
\includegraphics[angle=270,scale=0.70615]{IC1101_Maj_Min_fit.ps}

\vspace{-.05cm}
\hspace*{-.26055cm}
\includegraphics[angle=270,scale=0.70615]{IC1101_PA_B4.ps}
\vspace{-.25cm}
\caption{1D multi-component decompositions of the (i) {\it HST}
  $F702W$-band geometric mean ($\sqrt ab = $ $a \sqrt (1- \epsilon)$)
  surface brightness profile (left), (ii) {\it HST} $F702W$-band
  semi-major axis profile (middle) and (iii) composite semi-major axis
  profile (i.e., {\it HST} $F702W$-band profile at $R \la 40 \arcsec $
  plus SDSS $r$-band profile at $R > 40 \arcsec$) of IC 1101.  From
  top to bottom, the panels show Gaussian nuclear component (triple
  dot-dashed curve) + core-S\'ersic spheroid (dashed curve) + S\'ersic
  intermediate-component (dot-dashed curve) + exponential halo (dotted
  curve) model fits together with the rms residuals about the fits,
  ellipticity ($\epsilon$), position angle (P.A., measured in degrees
  from north to east) and isophote shape parameter ($B_{4}$) profiles
  for the galaxies. The models have been convolved with the PSF (see
  the text for details).}
\label{Fig5} 
 \end{figure*}

%\subsection{Fitting analysis}\label{Sec3.1}
 Figs.~\ref{Fig5} and \ref{FigMnG} show a four-component decomposition
 of the {\it HST} $F702W$ geometric mean radius ($\sqrt ab$) and
 semi-major axis light profiles of IC~1101 into a spheroid, an
 intermediate-scale component, an outer stellar halo and small
 Gaussian for the extended nucleus apparent in the ellipticity
 profile. The fit residuals and the root-mean-square (rms) residuals
 are also shown.  The best-fit parameters which match the data are
 calculated by iteratively minimizing the rms residuals using the
 Levenberg-Marquardt optimisation algorithm (see
 \citealt{2014MNRAS.444.2700D}; \citealt{2016MNRAS.462.3800D}). For
 each iteration, the profiles of individual model components were
 convolved with the Gaussian point-spread function (PSF) and then
 summed to create the final model profile. The ellipticity was taken
 into account when convolving the major-axis light profiles (see
 \citealt{2001MNRAS.321..269T, 2001MNRAS.328..977T}).
 The FWHMs of the PSFs were measured using several stars in the galaxy
 image. While we focus throughout this paper on the 1D fit to the {\it
   HST}
 $F702W$ light profiles convolved with the Gaussian PSF, we also
 performed a 1D fit to IC~1101's {\it HST}
 $F450W$ light profile and to the {\it HST}
 $F702W$ light profile convolved with the Moffat PSF to check the
 discrepancies in fit parameters arising from using the
 $F702W$- and
 $F450W$-band images, and our treatment of the PSF
 (Fig.~\ref{FigMnG}). We found good agreement between the fits to the
 $F702W$- and
 $F450W$-band profiles. Also, Figs.~\ref{Fig5} and \ref{FigMnG} reveal
 an excellent agreement between the fits convolved with the Gaussian
 and Moffat PSFs. We note that the {\it HST} WFPC2 PSF is better
 described by the Moffat function than the Gaussian function. However,
 our Gaussian and Moffat PSF- convolved fits agree very well because
 IC~1101 has a flat core. Tables~\ref{Tab2} and \ref{Tab3} list the best-fit
 model parameters from the 1D decompositions.

 The four-component (core-S\'ersic spheroid) plus (S\'ersic
 intermediate-scale component) plus (exponential halo) plus (Gaussian
 nuclear component) model yields an excellent fit to both the
 geometric mean profile and the major-axis profile of IC~1101,
 revealed by very small rms residuals of 0.016 mag
 arcsec$^{-2}$ and 0.013 mag
 arcsec$^{-2}$, respectively (Fig.~\ref{Fig5}). The initial fit to the
 light profile of IC~1101 using a core-S\'ersic model was inadequate,
 as the intermediate-scale and halo components caused obvious residual
 structures. Adding a halo component improved our fit, but from the
 fit and the residual profile it was clear that an additional
 intermediate-scale component and a small central Gaussian component
 are also present in the galaxy. Our adopted four-component
 decomposition (Fig.~\ref{Fig5}, left and middle panels) is supported
 by the galaxy colour, velocity dispersion profile and isophotal
 properties (see Sections \ref{Sec3.2} and \ref{Sec3.3}). The fit to
 the geometric mean radius gives a core-S\'ersic spheroid with a break
 radius of $R_{\rm b} \sim 2\farcs 52 \pm 0.07 \approx 3.8 \pm
 0.1$ kpc and a S\'ersic index of $n \sim
 6.3$, while the major-axis fit yielded $R_{\rm b} \sim 2\farcs 77 \pm
 0.07 \approx 4.2~\pm~0.1$ kpc and $n \sim
 5.6$.  IC~1101 has the largest core size ($R_{\rm b,maj} \sim
 4.2$ kpc) to date.
The ``intermediate-scale component'' has a low ($n \sim 0.62$)
S\'ersic stellar light distribution (Table~\ref{Tab2}). The outer halo
light of IC~1101 is well fit by an exponential ($n=1$) function (e.g.,
\citealt{2007MNRAS.378.1575S}; \citealt{2008A&A...483..727P}), and the
inner Gaussian component has an apparent $F702W$ magnitude of 22.8
mag.

Inside $R \sim 0\farcs
2$, the ellipticity of IC~1101 increases from $\epsilon \sim
0.1$ to $\epsilon \sim
0.4$ towards the centre, which seems to suggest the presence of a
nuclear disc (Fig.~\ref{Fig5}). Also, the isophotes of the galaxy are
discy (i.e., B$_{4} >$ 0) inside $R \sim0\farcs
1$. PSF tends to circularise the isophotes.  Using {\it HST} WFPC2
images, \citet{2001AJ....121.2431R} noted that the subpixel
interpolation routine of the IRAF task can underestimate the isophote
ellipticity and generate artificial deviations from pure ellipses at
$R \la 0\farcs
2$ (see also \citealt{1987MNRAS.226..747J}). However, as noted above
IC~1101 has isophotes with high ellipticities at $R \la 0\farcs
2$.  The faint nuclear component of IC~1101 may be due to an
unresolved double nucleus which is produced by a low-luminosity AGN
and the leftover stellar nuclei of an accreted satellite that was
tidally disrupted by the central SMBH. The elliptical galaxies
NGC~4486B, VCC 128 and NGC 5419 contain two central point sources
separated by 12 pc, 32 pc and 70 pc, respectively, that created a
double nucleus with high ellipticity and positive
$B_{4}$ values (see \citealt[their Figs.\ 3 and
4]{1996ApJ...471L..79L}; \citealt{2005AJ....129.2138L};
\citealt{2006ApJ...651L..97D}; \citealt{2012ApJ...755..163D};
\citealt{2016MNRAS.462.2847M}).  The NRAO VLA Sky Survey (NVSS,
\citealt{1998AJ....115.1693C}), with a resolution of
45$\arcsec$, has detected a radio source at 1.4 GHz with an integrated
flux of $527 \pm
18.2$ mJy. This source has an optimal location
$4\arcsec$ south-west of IC~1101's centre, favouring the presence of
an AGN in IC 1101. NVSS has also detected a weaker radio source with
an integrated flux of $5.3 \pm 0.5$ mJy located $\sim 50
\arcsec$ away from the stronger source. We cannot rule out the
possibility of a dual AGN near/at the centre of IC 1101.

As discussed in Section~\ref{Sky}, the {\it HST} $F702W$ and SDSS
$r$-band light profiles somewhat disagree at large radii
($R \ga 40\arcsec$) due to a poor sky background subtraction by the
SDSS pipeline (Fig.~\ref{Fig55}). We checked on the level of this
discrepancy between the profiles by fitting a four-component
(core-S\'ersic spheroid) plus (S\'ersic intermediate-scale component)
plus (exponential halo) plus (inner Gaussian) model to a composite
light profile, i.e., {\it HST} $F702W$ profile at $R \la 40\arcsec$
plus SDSS $r$-band data over $R > 40\arcsec$ (Fig.~\ref{Fig5}). This
fit agrees with the one done using the {\it HST} $F702W$ light profile
(see Fig.~\ref{Fig5} and Tables~\ref{Tab2} and \ref{Tab3}) and
therefore suggests that the sky-background subtraction is not a big
issue.

A key point highlighted by \citet{2012ApJ...755..163D,
  2013ApJ...768...36D} is the misclassification of (low-luminosity
$M_{B} \ga -20.5$ mag) coreless galaxies with low S\'ersic index
profiles as galaxies with depleted cores by the Nuker model.  A
PSF-convolved four-component S\'ersic spheroid + S\'ersic
intermediate-scale component + exponential halo + (inner) Gaussian fit
fails to describe the geometric-mean light profile of IC~1101
(Fig.~\ref{FigMnG}).  This fit yields a near exponential ($n=1.04$)
profile to approximate the core-S\'ersic spheroid profile, and an
intermediate S\'ersic model component with a half-light radius surface
brightness $\mu_{\rm e} \sim 22.3$ mag arcsec$^{-2}$, $\sim 0.5$ mag
arcsec$^{-2}$ brighter than that of our adopted geometric-mean fit,
$\mu_{\rm e} \sim 22.8$ mag arcsec$^{-2}$ (Table~\ref{Tab2}).  The
pattern in the residual profile between
$0 \farcs 2 \la R \la 3\arcsec$ reveals the presence of a depleted
core in the galaxy (Fig.~\ref{FigMnG}).

\subsection {2D decomposition}\label{2Deco}

We fit a 2D model, comprising a Gaussian nuclear component, a
core-S\'ersic spheroid, a S\'ersic intermediate-scale component and an
exponential stellar halo, to the WFPC2 $F702W$ image of IC~1101 using
{\sc imfit\footnote{http://www.mpe.mpg.de/~erwin/code/imfit/}}
\citep{2015ApJ...799..226E} and the mask image from the IRAF {\sc
  ellipse} run (Section~\ref{Surf_br}). Fig.~\ref{Fig3} (right) shows
this model image that was convolved with a Tiny Tim WFPC2 PSF
\citep{1995ASPC...77..349K}. Fig.~\ref{FigRes} (right) shows the
pertaining {\sc imfit} residual image created after subtracting the
model image from the galaxy image. In general, the agreement between
the 1D and 2D decompositions is good, but the $\mu_{\rm 0,h}$ of the
2D halo is brighter than that of the 1D one (see Figs.~\ref{Fig3}, and
\ref{FigRes}, and Tables~ \ref{Tab2}, \ref{Tab3} and
\ref{Tab33}). However, because each {\sc imfit} galaxy component is
associated with a single ellipticity, position angle and isophote
shape parameter, IC~1101's ellipticity gradient over $R \la 4\arcsec$
was not well represented by our best $\epsilon \sim 0.24$
core-S\'ersic model component that dominates other fit components at
$R \la 10\arcsec$, resulting in the residual structure (see
Figs.~\ref{Fig3}, \ref{FigRes} and \ref{Fig5}). \citet[their
Fig.~4]{2015ApJ...807..136B} also performed 2D
core-S\'ersic+exponential and S\'ersic+exponential fits to the 2D
light distribution of the BCG Holm 15A with outwardly rising
ellipticity using the {\sc galfit-corsair} software
\citep{2014PASP..126..935B} and found residual structures similar to
that shown in Fig.~\ref{FigRes} (right). \citet{2016PASA...33...62C}
presents a discussion regarding the pros and cons of how galaxy images
are modelled, including a discussion of the benefits of modelling 1D
light profiles. We prefer our 1D decompositions and focus on them
throughout the paper, but the good agreement between the 1D and 2D fit
parameters, except for $\mu_{\rm 0, h}$, implies that the main
conclusions in this paper remain unchanged if we choose our 2D
decomposition.

\subsection{Colour map} \label{Sec3.2}

\begin{figure*}
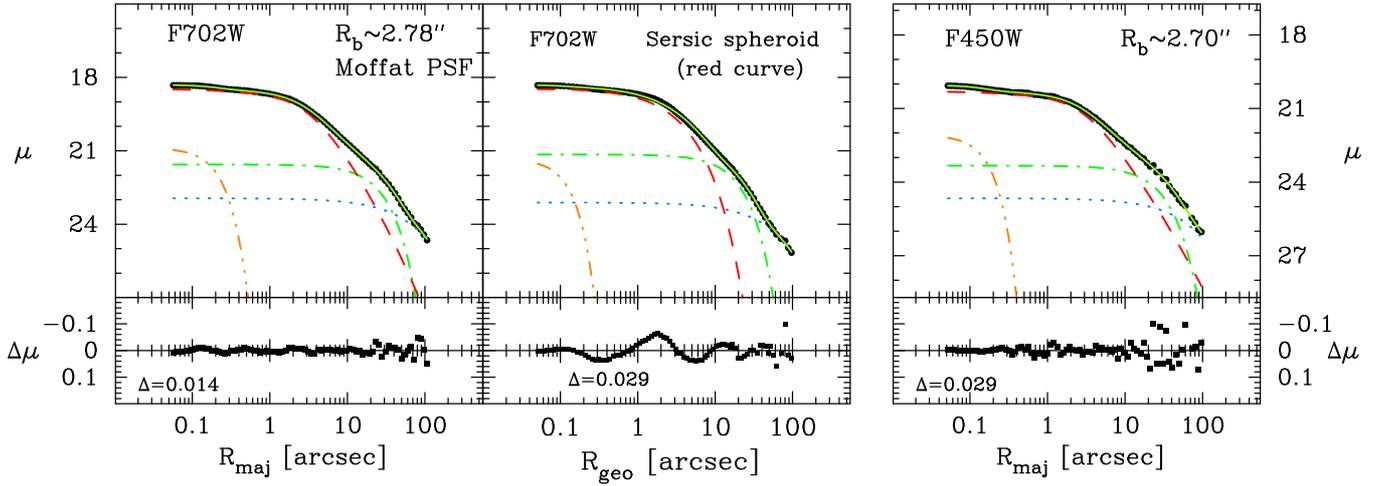

 \begin{tabular}{|c|c|c|}
\hspace*{-.40806453901270899cm}   
\includegraphics[angle=270,scale=0.646525]{Moffat_psf.ps}&
\hspace*{-.568453901270899cm}   
\includegraphics[angle=270,scale=0.646525]{Sersic_fit.ps}&
\hspace*{-.009951391568453901270899cm}   
\includegraphics[angle=270,scale=0.646525]{Blue_band_fit.ps}
\vspace*{-.32615446453901270899cm}   
\end{tabular}
\caption{ Left-hand panel: similar to Fig.~\ref{Fig5} (middle) but
  showing here a four-component fit convolved with a Moffat PSF. Middle panel: a
  Gaussian nuclear component (triple dot-dashed curve) + S\'ersic
  spheroid (dashed curve) + S\'ersic intermediate-component
  (dot-dashed curve) + exponential halo (dotted curve) fit to the
  $F702W$-band geometric mean surface brightness profile of
  IC~1101. Our fitting code attempts to fit a near exponential
  ($n\sim1$) S\'ersic model to what is actually a core-S\'ersic light
  distribution of the Spheroid with $n\sim 6.3$, yielding a snake like
  residual structure at $R \ga 0 \farcs 2$ and a brighter
  intermediate-scale component than that of our adopted fit (see
  Section~\ref{Sec3}). Right-hand panel: similar to Fig.~\ref{Fig5} (middle) but
  showing here a four-component fit to the WFPC2 $F450W$ light profile.}
\label{FigMnG} 
 \end{figure*}

\begin{figure}
 \hspace*{-.880199cm}   
\vspace*{-.4199cm}   
 \begin {minipage}{84mm}
\includegraphics[angle=0,scale=.314504]{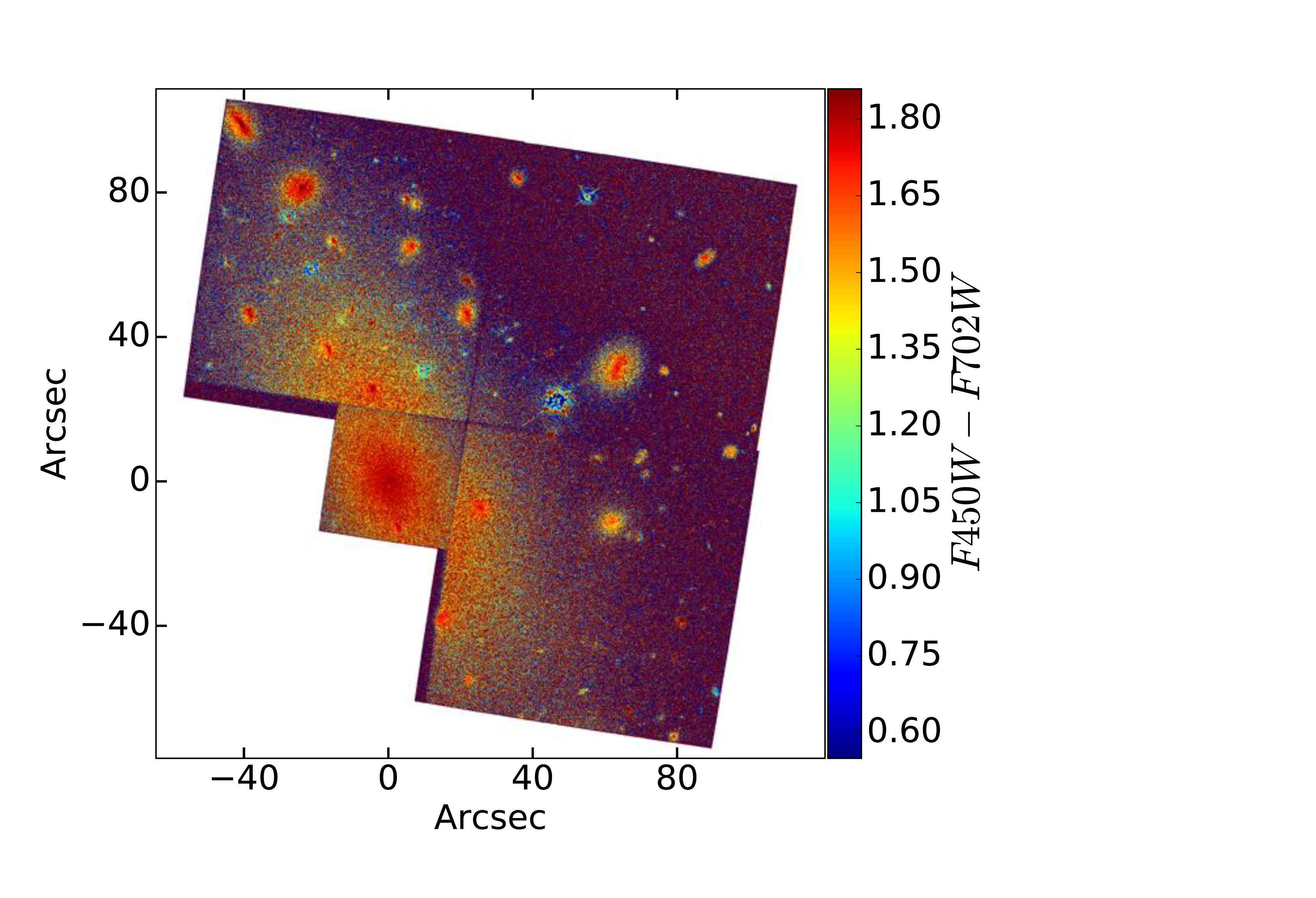}
\vspace{-1.050355cm}
 \caption{{\it HST} WFPC2 $F450W$ $-$ $F702W$ colour map of IC~1101
   shows that the galaxy becomes bluer with increasing radius, in good
   agreement with our brightness profile decomposition
   (Fig.~\ref{Fig5}). The regions where the intermediate-component and
   the halo light dominate are bluer than that dominated by the
   spheroid. }
\label{Fig7} 
\end {minipage}
\end{figure}

Fig.~\ref{Fig7} shows the WFPC2 $F450W - F702W$ colour map for
IC~1101, indicating that there is no obvious localised dust
absorption. To create this colour map, the WFPC2/$F450W$ and
WFPC2/$F702W$ images were aligned to better than 0$\farcs$2 using
stars in the images with the IRAF GEOMAP and GEOTRAN tasks. We did not
worry about the difference between the PSFs of the WFPC2/$F450W$ and
WFPC2/$F702W$ images. The galaxy becomes gradually bluer at larger
radii where the intermediate-scale component and the halo light
dominate compared to the inner region with a high concentration of old
stars from the spheroid. As noted in Section~\ref{Sec33.1}, we found
good agreement between the 1D $F450W$-band and $F702W$-band light
profile decompositions (Figs.~\ref{Fig5} and \ref{FigMnG}). The slight
discrepancies in the fit parameters are attributed to the galaxy
colour gradient.

\begin{table*}
\setlength{\tabcolsep}{0.0488240in}

\begin {minipage}{186mm}
~~~~~~~~~~~\caption{Structural parameters}
\label{Tab2}
\begin{tabular}{@{}lllcccccccccccccccccccccccccccccccccccccccccccccc@{}}
\hline
\hline
Galaxy&$ \mu_{\rm b} $ & $R _{\rm b}$ &$R_{\rm b}$ &$
                                                           \gamma$&$\alpha$&$n_{\rm
                                                                             cS}$&$R_{\rm
                                                                                   e,cS}$&$R_{\rm
                                                                                           e,cS}$&$\mu_{\rm
                                                                                                   e,S
                                                                                                   }
                                                                                                   $&$n_{S}$&$R_{\rm
                                                                                                              e,S}$&$R_{\rm
                                                                                                                     e,S}$&$\mu_{\rm
                                                                                                                            0,h}$&$h$&$h$&m$_{\rm
                                                                                                                                           pt}$\\
&&(arcsec)&(kpc)&&&&(arcsec)&(kpc)&&&(arcsec)&(kpc)&&(arcsec)&(kpc)&(mag)\\
(1)&(2)&(3)&(4)&(5)&(6)&(7)&(8)&(9)&(10)&(11)&(12)&(13)&(14)&(15)&(16)&(17)\\
\multicolumn{6}{c}{} \\ 
\hline
 &&&&&&&&{\it 1D fit/$HST~$F702W$$ }&&&&&&&  \\                              
 IC~1101        & 19.33  &2.52  &3.8 &0.05 &2   &6.32   &5.3  &8.0&22.81  &0.62&19.5&29.4& 23.41 &58.6&88.5&22.8\\
\hline
\hline
\end{tabular} 
Notes.--- Structural parameters from the 1D Gaussian nuclear component
+ core-S\'ersic spheroid + S\'ersic intermediate-component +
exponential halo model fit to the {\it HST}
$F702W$ geometric-mean-axis surface brightness profiles of IC~1101
(Fig.~\ref{Fig5}). The fit was convolved with a Gaussian PSF. Col.\ 1:
galaxy name. Cols.\
2$-$9: best-fit parameters from the core-S\'ersic model. Cols.\ $10-
13$: S\'ersic model parameters for the intermediate-scale
component. Cols.\
14$-$16 best-fit parameters for the exponential halo light. Col.\ 17:
apparent $F702W$ magnitude of the unresolved double nucleus. $
\mu_{\rm b}$, $ \mu_{\rm e}$ and $\mu_{0,\rm
  h}$ are in mag
arcsec$^{-2}$. The core-S\'ersic spheroid has a total
$F702W$-band absolute magnitude $M_{\rm Sph} \sim$
$-$ 24.3 mag. The intermediate-component has a total
$F702W$-band absolute magnitude $M_{\rm Int} \sim$
$-$ 23.9 mag.  These magnitudes are
$not$ corrected for Galactic dust extinction,
(1+z)$^{4}$ surface brightness dimming, and stellar evolution from $z
\sim$ 0.08 to the present day. We estimate that the uncertainties on
the core-S\'ersic parameters $R_{\rm b}$, $\gamma$, $n_{\rm
  cS}$ and $R_{\rm e,cS}$ are
$\sim$ 2.5\%, 10\%, 20\% and 25\%, while the uncertainties associated
with the S\'ersic parameters $n_{\rm S}$, $R_{\rm
  e,S}$ and the exponential scale length $h$ are
$\sim$ 20\%, 25\% and 10\%, respectively. The uncertainties on
$\mu_{\rm b}$, $\mu_{\rm e,S}$, and $\mu_{\rm 0,h}$ are
$\sim$ 0.02 mag arcsec$^{-2}$, 0.2 mag
arcsec$^{-2}$ and 0.2 mag
arcsec$^{-2}$, respectively. These errors on the fit parameters were
estimated following the techniques in \citet[their
Section~4]{2012ApJ...755..163D}.
\end{minipage}
\end{table*}

\begin{landscape}
\begin{table}
\setlength{\tabcolsep}{0.0546055173190672430in}
\begin {minipage}{225mm}
~~~~~~~~~~~\caption{Structural parameters}
\label{Tab3}
\begin{tabular}{@{}lllccccccccccccccccccccccccccccccccccccccccccccc@{}}
  \hline
  \hline
  Galaxy&$ \mu_{\rm b} $ & $R _{\rm b}$ &$R_{\rm b}$ &$
                                                             \gamma$&$\alpha$&$n_{\rm
                                                                               cS}$&$R_{\rm
                                                                                     e,cS}$&$R_{\rm
                                                                                             e,cS}$&$\mu_{\rm
                                                                                                     e,S
                                                                                                     }
                                                                                                     $&$n_{S}$&$R_{\rm
                                                                                                                e,S}$&$R_{\rm
                                                                                                                       e,S}$&$\mu_{\rm
                                                                                                                              0,h}$&$h$&$h$&m$_{\rm
                                                                                                                                             pt}$\\
  &&(arcsec)&(kpc)&&&&(arcsec)&(kpc)&&&(arcsec)&(kpc)&&(arcsec)&(kpc)&(mag)\\
  (1)&(2)&(3)&(4)&(5)&(6)&(7)&(8)&(9)&(10)&(11)&(12)&(13)&(14)&(15)&(16)&(17)&\\
  \multicolumn{6}{c}{} \\ 
  \hline
&&&&&&&&{\it 1D fit/$HST~$F702W$$ }&&&&&&&  \\ 
IC~1101        & 19.35  &2.77 &4.2 &0.08 &2   & 5.60  &7.7 &11.6& 22.68~& 0.67&26.2&39.5& 22.94 &66.70&100.7&23.1\\
&&&&&&&&{\it 1D fit/$HST~ $F702W$$ + SDSS~$r$}& &&&&&&&&&&&& \\
IC~1101         & 19.33  &2.76 &4.2 &0.08&2  & 6.29 &9.8&14.8& 22.74& 0.58&24.1&36.4& 22.80 &69.24&104.6&23.1\\
&&&&&&&&{\it 1D fit/$HST~ $F702W$$  (Moffat PSF)}& &&&&&&&&&&&& \\
IC~1101       & 19.33  &2.78 &4.2 &0.08&2  & 5.70 &7.7&11.6& 22.67& 0.65&26.2&39.5& 22.94 &67.12&101.4&23.0\\
&&&&&&&&{\it 1D fit/$HST~ $F450W$$  } & &&&&&&&&&&&& \\
IC~1101         &21.14  &2.65 &4.0 &0.07&2  & 7.50 &13.1&19.8&24.52& 0.71&30.0&45.3& 24.66 &66.83&100.9&24.8\\
\hline
 \hline
\end{tabular} 

Notes.--- Same as Table~\ref{Tab2} but here showing semi-major axis structural parameters
from the fits to the $HST~F702W$, $HST~F450W$ and   composite ({\it HST~F702W} plus SDSS
$r$-band) profiles of IC~1101. We also provide fit parameters from the
1D fit to the $HST~F702W$ light profile of the galaxy convolved with a Moffat PSF. 
\end{minipage}
\end{table}
\end{landscape}

\begin{landscape}
\begin{table}
\setlength{\tabcolsep}{0.044046055173190672430in}
\begin {minipage}{233mm}
\caption{Structural parameters}
\label{Tab33}

\begin{tabular}{@{}lllccccccccccccccccccccccccccccccccccccccccccccc@{}}
  \hline
  \hline
  Galaxy&$ \mu_{\rm b} $ & $R _{\rm b}$ &$R_{\rm b}$ &$
                                                             \gamma$&$\alpha$&$n_{\rm
                                                                               cS}$&$R_{\rm
                                                                                     e,cS}$&$R_{\rm
                                                                                             e,cS}$&P.A.$_{cS}$&$\epsilon_{cS}$&$\mu_{\rm
                                                                                                     e,S
                                                                                                     }
                                                                                                     $&$n_{S}$&$R_{\rm
                                                                                                                e,S}$&$R_{\rm
                                                                                                                       e,S}$&P.A.$_{S}$&$\epsilon_{S}$&$\mu_{\rm
                                                                                                                              0,h}$&$h$&$h$&P.A.$_{\rm
                                                                                                                                             h}$&$\epsilon_{\rm
                                                                                                                                                  h}$&m$_{\rm
                                                                                                                                             pt}$\\
  &&(arcsec)&(kpc)&&&&(arcsec)&(kpc)&($^{\circ}$)&&&&(arcsec)&(kpc)&($^{\circ}$)&&&(arcsec)&(kpc)&($^{\circ}$)&&(mag)\\
  (1)&(2)&(3)&(4)&(5)&(6)&(7)&(8)&(9)&(10)&(11)&(12)&(13)&(14)&(15)&(16)&(17)&\\
  \multicolumn{6}{c}{} \\ 
  \hline
{\it 2D fit/$HST~$F702W$$} \\
IC~1101 &19.39  &2.79&4.4 &0.14&2.4&6.4 & 5.4 &8.2&18.4&0.24&22.29&0.65&19.1&28.8&21.1& 0.57 &22.24&57.7&87.1&38&0.50&22.91\\
\hline
 \hline
\end{tabular}  

Notes.--- Similar to Table~\ref{Tab2} but here showing structural
parameters from the 2D {\sc imfit} (nuclear component + core-S\'ersic
spheroid + S\'ersic intermediate-component + exponential halo) model
image fit to the
$HST~F702W$-band image. The {\sc imfit} isophote shape parameter
$c_{0,S}$, not to be confused with that of the IRAF/{\sc ellipse}
$B_{4}$, is positive/negative for boxy/discy isophotes
\citep{2015ApJ...799..226E}. We found $c_{0,S} \sim 0.61 \pm
0.007$ and $c_{\rm 0,h} \sim 0.79 \pm
0.007$ for the intermediate-component and outer halo. The
uncertainties from {\sc imfit} associated with $R_{\rm b}$,
$\gamma$, $\alpha$, $n_{\rm cS}$, $R_{\rm e,cS}$,
P.A.$_{cS}$, $\epsilon_{cS}$, $n_{\rm S}$, $R_{\rm e,S}$,
P.A.$_{S}$, $\epsilon_{S}$, $h$, P.A.$_{\rm h}$, $\epsilon_{\rm
  h}$ are
$\sim$ 0.6\%, 0.3\%, 0.3\%, 0.4\%, 0.8\%, 0.3\%, 0.3\%, 0.3\%, 0.9\%,
0.3\%, 0.3\%, 0.4\%, 0.3\% and 0.3\%. The uncertainties on $\mu_{\rm
  b}$, $\mu_{\rm e,S}$, and $\mu_{\rm 0,h}$ are
$\sim$ 0.01 mag arcsec$^{-2}$, 0.01 mag
arcsec$^{-2}$ and 0.02 mag arcsec$^{-2}$, respectively.
 \end {minipage}
\end{table}
%\end{sidewaystable}
\end{landscape}

\subsection{Ellipticity, position angle and isophote disciness/boxiness}\label{Sec3.3}

Excluding the most PSF-affected region (i.e., $R \la 0\farcs2$, see
Section \ref{Sec3}), IC~1101 has an ellipticity of 0.1 $\pm$ 0.08
within the break radius $R_{\rm b}=2\farcs 52$, in good agreement with
the observations by \citet{2015ApJ...798...55D} who showed that
core-S\'ersic galaxies tend to have round cores, i.e., $\epsilon$
($R=R_{\rm b}$) $\la 0.2$.  The ellipticity of IC~1101 increases
steadily from $\sim$ 0.2 to 0.5 with increasing radius over
$3\arcsec \la R_{\rm geo} \la 20\arcsec$, it remains roughly constant
(i.e., $\epsilon \sim 0.5$) at radii dominated by the
intermediate-scale component
($20\arcsec \la R_{\rm geo} \la 40\arcsec$) and then increases again
at large radii ($R_{\rm geo} \ga 40\arcsec$) where the halo light
dominates (Figs.~\ref{Fig4},~\ref{Fig55} and \ref{Fig5}). The trend of
increasing ellipticity with radius has been observed for other BCGs
(e.g., \citealt{1991AJ....101.1561P}; \citealt{1995ApJ...440...28P};
\citealt[their Fig.~10]{2005ApJ...618..195G};
\citealt{2006MNRAS.372L..68K}; \citealt[their
Fig.~7]{2015ApJ...807...56B}).

The major axis of IC~1101 is oriented in the south-west to north-east
direction (Figs.~\ref{Fig3}, \ref{Fig4} and \ref{Fig5}). The spheroid
and the intermediate-scale component are very well aligned with only a
modest twist of $\sim 5^{\circ}$. The outer halo is twisted by
$\sim 20^{\circ}$ from the intermediate-scale component. The shape and
orientation of the halo of IC~1101 are comparable to those of the
surrounding cluster galaxies, suggesting a strong connection between
the cluster galaxies and the host cluster's gravitational potential
(\citealt{1978ApJ...226...55D, 1978ApJ...223..765D,
  1979ApJ...231..659D}).

The isophotes of IC~1101 are predominantly boxy (i.e., $B_{4} < 0$) at
$0\farcs1 \la R_{\rm geo} \la 1 \arcsec$ and
$R_{\rm geo} \ga 20\arcsec$ (Figs.~~\ref{Fig1}, \ref{Fig4} and \ref{Fig5}). The
galaxy has  pure elliptical (i.e., $B_{4} = 0$) isophotes over
$1\arcsec \la  R_{\rm geo} \la 20\arcsec$.

\section{Discussion}\label{Sec4}
 \subsection{Past work on IC~1101 and similar BCGs}\label{Sec4.0}
 
 BCGs have been shown to contain excess halo light at large radii with
 respect to the \citet{1948AnAp...11..247D} $R^{1/4}$ model fit to the
 main body of the BCG that is embedded within the halo light (e.g.,
 \citealt{1976ApJ...209..693O}; \citealt{1977MNRAS.178..137C};
 \citealt{1981ApJ...243...26D}; \citealt{1984ApJ...286..106L};
 \citealt{1986ApJS...60..603S}). However, a single-component $R^{1/n}$
 galaxy will ``appear'' to have a halo if $n > 4$.  \footnote{While
   \citet{2005ApJ...618..195G} advocated an $R^{1/4}$ spheroid plus an
   $R^{1/4}$ outer stellar halo model to describe the profile of their
   BCGs, \citet{2003Ap&SS.285...67G} fit a two-component
   double-S\'ersic model to the light profiles of
   BCGs.}\citet{2005ApJ...618..195G} and \citet{2005MNRAS.358..949Z}
 fit an $R^{1/4}$ spheroid plus an $R^{1/4}$ outer halo light model to
 the light profiles of their BCGs before \citet{2007MNRAS.378.1575S}
 showed that most BCGs in their sample are well described by a
 S\'ersic ($R^{1/n}$) spheroid plus an exponential ($n=1$) halo model,
 excluding the core regions (see also \citealt
 {2008A&A...483..727P}). \citet{2011ApJS..195...15D} also found that
 roughly half of their 430 BCGs are well fit by a S\'ersic spheroid
 plus an outer exponential model.

 IC 1101 and its host cluster (A2029) have been extensively studied
 (e.g., \citealt{1964ApJ...140...35M}; \citealt{1978ApJ...226...55D,
   1978ApJ...223..765D, 1979ApJ...231..659D};
 \citealt{1984ApJ...285....1S}; \citealt{1987ApJS...64..643S};
 \citealt{1996AJ....112..797S}; \citealt{1991AJ....101.1561P};
 \citealt{1991ApJ...369...46U}; \citealt{1999MNRAS.307..131C};
 \citealt{2002ApJ...576..720K}). As noted in Section~\ref{Sec3},
 IC~1101 has an intermediate-scale component plus an exponential
 stellar halo that dominate over the spheroid's core-S\'ersic stellar
 light distribution at large radii.  Computing the total integrated
 flux for each model component (Fig.~\ref{Fig5}) using the best-fit
 structural parameters (Table~\ref{Tab2}) yields total light fractions
 for the spheroid, intermediate-scale component and outer halo
 component equal to 25.1\%, 17.4\% and 57.5\%, respectively. The total
 flux fraction of the unresolved double nucleus is negligible.

 The intermediate-scale component plus the outer stellar halo and the
 outwardly rising ellipticity profile (Fig.~\ref{Fig5} and
 Section~\ref{Sec3.3}) are the likely reasons why IC~1101 is
 classified as a lenticular galaxy in the RC3
 (\citealt{1991rc3..book.....D}). IC~1101 shows almost no rotation
 ($V_{\rm rot} \sim$ 0 km s$^{-1}$) within 13 kpc (\citealt[his
 Fig.~4b]{1979ApJ...231..659D}; \citealt[their
 Fig.~5b]{1995ApJ...438..539F}). At $R \ga 13$ kpc, \citet[his
 Fig.~4b]{1979ApJ...231..659D}, measured $V_{\rm rot} $
 $\sim 50 - 150$ km s$^{-1}$ but he warned that those results were
 tentative. Importantly, \citet[his Fig.~5]{1979ApJ...231..659D}
 showed that the velocity dispersion ($\sigma$) of IC 1101 increases
 outward from 375 km s$^{-1}$ within $R \sim 5$ kpc to $\sim$ 425
 $\pm$ 25 km s$^{-1}$ for 6 kpc $\la R\la 62$ kpc before rising to
 $\ga$ 500 km s$^{-1}$ beyond $R \sim 62$ kpc (see also
 \citealt{1985MNRAS.212..471C}; \citealt[their
 Fig.~6]{1996AJ....112..797S}). Not only is this $\sigma(R)$ profile
 consistent with our multi-component light profile decomposition, the
 low $V_{\rm rot}/\sigma$ ratio also indicates the absence of a disc
 confirmed by the boxy isophotes ($B_{4} < 0$) at $R \ga 20\arcsec$.
Akin to IC~1101, NGC 6166 (A2199-BCG) which possesses the largest
 core ($r_{\gamma'=0.5} \sim$ 1.5 kpc) in the
 \citet{2007ApJ...662..808L} sample of 219 galaxies also has outwardly
 rising velocity dispersion and ellipticity profiles (e.g.,
 \citealt{1995ApJ...438..539F}; \citealt{1999MNRAS.307..131C};
 \citealt{2002ApJ...576..720K}; \citealt{2015ApJ...807...56B}). The
 reason for a radially increasing velocity dispersion profile is the
 greater contribution from halo stars, which traces the cluster
 potential rather than the BCG potential.

 While the de Vaucouleurs $R^{1/4}$ model fit by \citet[his
 Fig.~2c]{1987ApJS...64..643S} to the ground-based light profile of
 IC~1101 was limited, the downward departure of the inner light
 profile from the model reveals the partially-depleted
 core. \citet{1991AJ....101.1561P} and \citet{2011ApJS..195...15D}
 claimed that IC~1101 does not have excess halo light, but see
 \footnote{\citet{1979ApJ...231..659D} interpreted the luminosity,
   mass distribution and $\sigma(R)$ profile of IC~1101 using a
   three-component isotropic King model to account for the main body
   of the BCG, the stellar halo of accreted luminous galaxies and dark
   matter which was not included in the luminosity distribution
   model. The double nucleus that we modelled (Fig~\ref{Fig5}) was not
   resolved in the \citet{1979ApJ...231..659D}
   image.}\citet{1979ApJ...231..659D} and
 \citet{1991ApJ...369...46U}. \citet{2011ApJS..195...15D} fit a single
 S\'ersic model to the light profile of IC~1101, yielding an extremely
 large half-light radius $R_{\rm e} = 439$ kpc and $n \sim 5.8$,
 although they adopted a two-component S\'ersic ($R^{1/n}$) plus
 exponential fit for their BCGs with $n \ga 8$ and $R_{\rm e} \ga$ 300
 kpc.  The \citet{2011ApJS..195...15D} ground-based data with a seeing
 of FWHM = $1\arcsec - 2 \arcsec$ did not have sufficient resolution
 to adequately resolve the core of IC~1101.

\begin{figure*}
 \hspace*{.05999cm}   
 \begin {minipage}{175mm}
~~~~~~~~~~~~~~~~~~~~~~~~~~~~~\includegraphics[angle=270,scale=0.608]{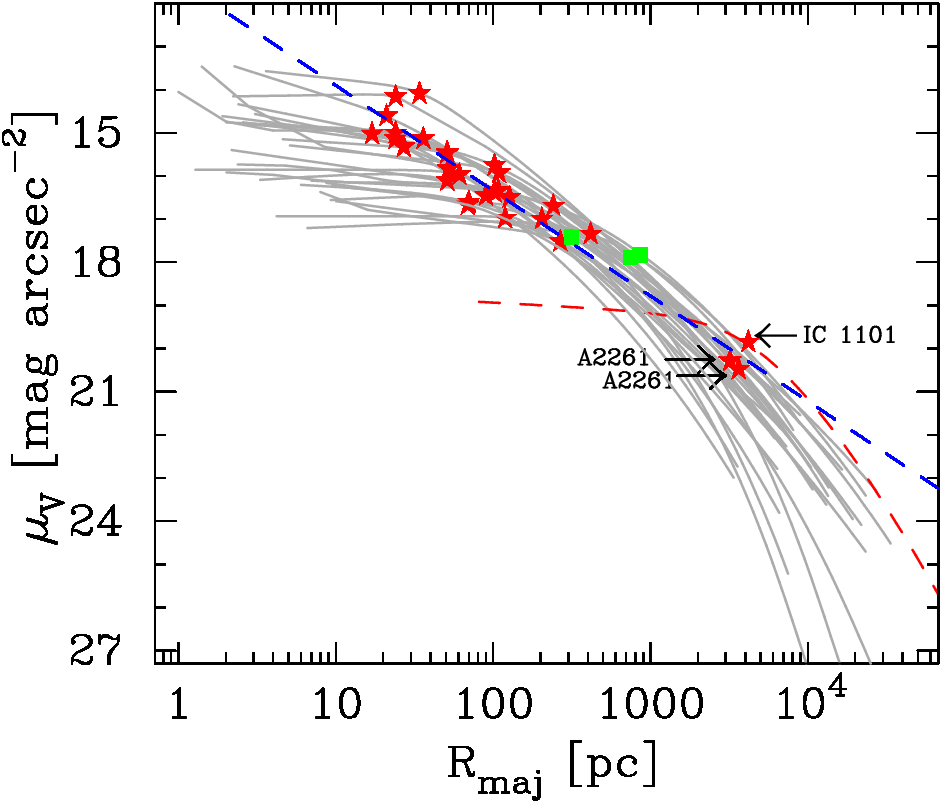} 
\caption{Compilation of core-S\'ersic fits to the major-axis surface
  brightness profiles of all core galaxies from \citet[their
  Table~2]{2014MNRAS.444.2700D} together with the $V$-band
  core-S\'ersic profile of IC~1101 from this work (red dashed
  curve). Red stars mark the break radius of the individual
  profiles. Green squares indicate the three galaxies (NGC 1600,
  \citealt{2016Natur.532..340T}; NGC 3842,
  \citealt{2014MNRAS.444.2700D}; NGC 4889, Dullo et al.\ in prep.)
  which have directly measured supermassive black hole masses
  $M_{\rm BH} \ga 10^{10} M_{\sun}$ (\citealt{2011Natur.480..215M};
  \citealt{2016Natur.532..340T}). For IC~1101, we show the major-axis
  core surface brightness ($\mu_{\rm b}$, Table~~\ref{Tab3}),
  corrected for Galactic dust extinction, (1+$z$)$^{4}$ surface
  brightness dimming, and stellar evolution using the MILES stellar
  library (\citealt{2010MNRAS.404.1639V}). For comparison, we include
  the hitherto largest core measured for A2261-BCG by
  \citet[$r_{\gamma=0.5} \sim$ 3.2 kpc]{2012ApJ...756..159P} and
  \citet[$R_{\rm b} \sim$ 3.6 kpc]{2016ApJ...829...81B}. IC~1101 and
  A2261-BCG follow the narrow sequence defined by the
  \citet{2014MNRAS.444.2700D} core-S\'ersic galaxies (see
  \citealt[their Fig.~8]{1997AJ....114.1771F}). The (blue) dashed line
  is a least-squares fit taken from \citet[their Table
  3]{2014MNRAS.444.2700D}.}
\label{Fig8} 
\end {minipage}
\end{figure*}

\subsection {The large partially-depleted core of IC~1101}\label{Sec4.1}

Fig.~\ref{Fig8} shows a compilation of major-axis core-S\'ersic model
profiles taken from \citet[their Table 2]{2014MNRAS.444.2700D}
together with that for IC~1101. IC~1101 has one of the faintest
$V$-band core (at $R_{\rm b}$ not to be confused with $R=0$) surface
brightnesses, $\mu_{\rm b} \sim$ 19.83 mag arcsec$^{-2}$, corrected
for Galactic dust extinction ($\sim -0.09$ mag arcsec$^{-2}$),
(1+$z$)$^{4}$ surface brightness dimming ($\sim -0.33$ mag
arcsec$^{-2}$), and stellar evolution from $z \sim$ 0.08 to the
present day, derived using the MILES stellar
library\footnote{http://miles.iac.es/pages/stellar-libraries/miles-library.php.}(\citealt{2010MNRAS.404.1639V}),
of $\sim +0.12$ mag arcsec$^{-2}$. For comparison, in addition to
Fig.~\ref{Fig8}, see \citet{2007ApJ...662..808L},
\citet{2012ApJ...756..159P}, \citet{2013AJ....146..160R} and
\citet{2014MNRAS.444.2700D}.

Although the break radius of IC~1101 ($R_{\rm b} \sim 4.2 \pm 0.1$
kpc) is roughly an order of magnitude larger than those typically
observed in core-S\'ersic early-type galaxies, it closely follows the
extrapolation of the $R_{\rm b} - \mu_{\rm b}$ sequence defined by
other core-S\'ersic galaxies (\citealt[their
Table~3]{2014MNRAS.444.2700D}; Fig.~\ref{Fig8}). The next-largest is
$ \sim 3.6$ kpc in the A2261 BCG (\citealt{2012ApJ...756..159P};
\citealt{2016ApJ...829...81B}) which is at redshift $z=0.2248$,
compared to $z=0.0799$ for IC~1101. For comparison, the three
core-S\'ersic galaxies (NGC~1600, NGC~3842 and NGC~4889) housing very
massive BHs ($M_{\rm BH}\ga 10^{10} M_{\sun}$) have break radii
$R_{\rm b} \sim 300$ pc $-$ 750~pc (\citealt{2014MNRAS.444.2700D},
$R_{\rm b,NGC~3842} \sim 315$~pc; \citealt{2016Natur.532..340T},
$R_{\rm b,NGC~1600} \sim 746$~pc; Dullo et al.\ in prep.,
$R_{\rm b,NGC~4889} \sim 861$~pc).

Using the best-fit parameters (Table~\ref{Tab2}), we find an absolute
magnitude $M_{\rm F702W} \approx$ $-$24.3 mag for the spheroid of
IC~1101. Correcting this for extinction, surface brightness dimming
plus stellar evolution and using
$V$$-$$F702W$ = 0.8 (\citealt{1995PASP..107..945F}) yields
$M_{V} \approx -23.8$ mag. This spheroid luminosity is very faint for
the galaxy's large break radius, as the major-axis $R_{\rm b}-M_{V}$
relation for core-S\'ersic galaxies (\citealt[their
Table~3]{2014MNRAS.444.2700D}) predicts a small core
($R_{\rm b,maj} \sim$ 400 pc) for $M_{V} \approx -23.8$ mag, compared
to IC~1101's $R_{\rm b,maj} \sim$ 4.2 kpc (Table~\ref{Tab3}).  Given
the 0.3 dex rms scatter ($\sigma_{s}$) for the $R_{\rm b}-M_{V}$
relation in the log ($R_{\rm b}$) direction, the break radius of
IC~1101 is 1.02 dex ($\sim$ 3.4$\sigma_{s}$) above the
'$R_{\rm b}-M_{V}$ relation. Considering the spheroid +
intermediate-scale component luminosity, the $R_{\rm b}-M_{V}$
relation (\citealt{2014MNRAS.444.2700D}) predicts $R_{\rm b,maj} \sim$
720 pc, i.e., IC~1101's break radius is 0.77 dex ($\sim$ 2.6
$\sigma_{s}$) above the $R_{\rm b}-M_{V}$ relation.

\subsection{Central stellar mass deficit ($M_{\rm def}$) of IC~1101}\label{Sec4.2}

In order to determine the stellar mass deficit of IC~1101, we follow
the prescription in \citet[see also \citealt{2004ApJ...613L..33G} and
\citealt{2006ApJS..164..334F} who adopted a slightly different
approach]{2014MNRAS.444.2700D}. We compute the difference in
luminosity between the inwardly-extrapolated outer S\'ersic profile of
the complete core-S\'ersic model (Eq.~\ref{Eqq2}) and the
core-S\'ersic model (Eq.~\ref{Eq2}). This yields a luminosity deficit
log~($L_{\rm def}/L_{\sun}$) $\sim$ 11.04$^{+0.07}_{-0.07}$ in the $F702W$-band which
corresponds to $\approx$ 27.7\% of the spheroid's luminosity prior to
the core depletion. For comparison, $L_{\rm def}$ is $\sim 58$\% of
the luminosity of the intermediate-scale component. Moreover,
\citet{2010MNRAS.407..447H} determined $L_{\rm def}$ using
model-independent analysis of galaxy light profiles. They found values
of  $L_{\rm def}$ consistent with those of e.g.,
\citet{2004ApJ...613L..33G}, \citet{2006ApJS..164..334F} and
\citet{2014MNRAS.444.2700D}.

The depleted core has an 
$F450W$$-$$F702W$ colour of 1.81 $\pm~0.03$ (Fig.~\ref{Fig7}), which
corresponds to a $V-I$ colour of 1.34 $\pm~0.03$
(\citealt{1995PASP..107..945F}). Using the ($V-I$)
colour-age-metallicity-$M/L_{V}$ diagram
(\citealt{2009MNRAS.397.2148G}) and assuming an old (12 Gyr) stellar
population for the core yields $M/L_{V} \sim 5.2$. This implies
$M/L_{F702W}$ $\sim$ 4.5 (\citealt[his Table 5A]{1994ApJS...95..107W})
and thus a stellar mass deficit ($M_{\rm
  def}$)~$\sim 4.9^{+1.03}_{-1.03} \times 10^{11} M_{\sun}$ in
IC~1101. The quoted 21\% uncertainty on $M_{\rm def}$
is based on the errors on the fit parameters (see Table~\ref{Tab2}).

There is no evidence for the presence of dust lanes in the central
regions of IC~1101 (see Figs.~\ref{Fig3}, \ref{FigRes}, \ref{Fig4} and
\ref{Fig7}) which could mimic a large apparent core/stellar mass
deficit. Comparing fits to optical and near-IR light profiles,
\citet[their Fig.~4]{2001AJ....122..653R} noted that nuclear dust
lanes in galaxies typically tend to flatten the inner negative
logarithmic slope ($\gamma$) of the profiles. For IC~1101, we find
that increasing $\gamma$ from 0.05 (Table~\ref{Tab3}) to 0.4 reduces
the stellar mass deficit of IC~1101 only by $\sim$ 20\%.

\citet{2006ApJ...648..976M} simulated the evolution of SMBH binaries that are
formed in galaxy merger remnants to quantify the relation between
merger histories of core-S\'ersic galaxies and their central stellar
mass deficits. His simulations first revealed that the accumulated
stellar mass deficit after $\mathcal{N}$ numbers of successive ``dry''
major merges is $M_{\rm def} \approx 0.5 \mathcal{N} M_{\rm BH}$,
where $M_{\rm BH}$ is the final mass of the SMBH.

IC~1101 does not have a directly determined SMBH mass. Therefore, we
estimate the SMBH mass using the $M_{\rm BH}-\sigma$ relation
(\citealt{2000ApJ...539L...9F}; \citealt{2000ApJ...539L..13G}), the
$M_{\rm BH}-L$ relation and $M_{\rm BH}- M_{*}$ relation
(\citealt{1989IAUS..134..217D}; \citealt{1995ARA&A..33..581K};
\citealt{1998AJ....115.2285M}).  For $\sigma \sim 378$ km s$^{-1}$
(HyperLeda), the \citet{2013ApJ...764..151G} ‘non-barred
$M_{\rm BH}-\sigma$ relation predicts
log~$M_{\rm BH}/M_{\sun} \sim 9.68 \pm 0.47$. The
\citet{2013ApJ...764..184M} $M-\sigma$ relation for early-type
galaxies including the two BCGs with very massive black holes (NGC
3842, $M_{\rm BH}/M_{\sun} \sim 9.7^{+3.0}_{-2.5} \times 10^{9}$ and
NGC 4889,
$M_{\rm BH}/M_{\sun} \sim 2.1 ^{+1.6}_{-1.6} \times 10^{10}$) and the
\citet[their Table 11]{2016ApJ...818...47S} $M-\sigma$ relation for
core-S\'ersic elliptical galaxies predict
log~$M_{\rm BH}/M_{\sun} \sim 9.63 \pm 0.50$ and
log~$M_{\rm BH}/M_{\sun} \sim 9.82 \pm 0.49$, respectively. Using the
$V$-band absolute magnitude of IC~1101's spheroid
($M_{\rm V} \approx -23.8$ mag, i.e.,
$M_{*} \approx 1.5 \times 10^{12} M_{\sun}$) and $B-V$=1.0
\citep{1995PASP..107..945F}, the \citet{2013ApJ...764..151G}
$M_{\rm BH}-L$ relation for core-S\'ersic galaxies gives
log~$M_{\rm BH}/M_{\sun} \sim 9.99 \pm 0.40$ and the
\citet{2016ApJ...817...21S} $M_{\rm BH}- M_{*}$ relation and the
\citet[their Table 11]{2016ApJ...818...47S} $M_{\rm BH}- M_{\rm Bu}$
relation for core-S\'ersic ellipticals yield
log~$M_{\rm BH}/M_{\sun} \sim 10.11 \pm 0.47$ and
log~$M_{\rm BH}/M_{\sun} \sim 9.68 \pm 0.45$, respectively.  Because
$M_{\rm Bu}$ is the total, rather than the stellar, mass of the bulge
\citep{2016ApJ...818...47S}, the value of
log~$M_{\rm BH}/M_{\sun} \sim 9.68 \pm 0.45$ derived here using
IC~1101's $M_{*}$ and the $M_{\rm BH}- M_{\rm Bu}$ relation is a lower
limit.

It follows that $M_{\rm def}/M_{\rm BH} \approx$ 38, 50 and 73 $-$ 101
for SMBH masses determined using the $M_{\rm BH}- M_{*}$,
$M_{\rm BH}-L$ and $M-\sigma$ relations, respectively. This translates
to spheroid formation via an extremely frequent ($\mathcal{N} \ga 76$)
major mergers (\citealt{2006ApJ...648..976M}), if nothing else was
occurring. Observations counting close galaxy pairs have led to
estimates that today's massive galaxies may have experienced 0.5 to 6
major mergers since $z \sim 3$ (e.g., \citealt{2006ApJ...640..241B};
\citealt{2007IAUS..235..381C}; \citealt{2012ApJ...747...34B};
\citealt{2012ApJ...747...85X}; \citealt{2014MNRAS.445.1157C}),
consistent with numbers estimated by theoretical models of galaxy
mergers involving SMBHs (\citealt[their
Fig.~2]{2002MNRAS.336L..61H}). The $M_{\rm def}/M_{\rm BH}$ ratio for
IC~1101 is an order of magnitude higher than the typical
$M_{\rm def}/M_{\rm BH}$ values published in the literature
(\citealt{2004ApJ...613L..33G}, $M_{\rm def}/M_{\rm BH} \sim 1- 2$;
\citealt{2006ApJS..164..334F}, mean $M_{\rm def}/M_{\rm BH} \sim 2.4$;
\citealt{2008MNRAS.391.1559H}, mean $M_{\rm def}/M_{\rm BH} \sim 2.3$;
\citealt{2013AJ....146..160R}, median
$M_{\rm def}/M_{\rm BH} \sim 2.2$ and $M_{\rm def}/M_{\rm BH}$
typically 0.2$-$10; \citealt{2013ApJ...768...36D,
  2014MNRAS.444.2700D}, $M_{\rm def}/M_{\rm BH} \sim 0.5 - 4$). In
Section~\ref{Sec4.3}, we discuss that if the depleted core of IC~1101
was created by binary SMBHs with final merged mass
$M_{\rm BH} > 10^{10} M_{\rm \sun}$ and enhanced core scouring
occurred due to a gravitational radiation-recoiled black hole
(\citealt{2008ApJ...678..780G}), then the number of major mergers
($\mathcal{N}$) that the galaxy underwent would be $\la 10$, in close
agreement with observations and theories.

\subsection{Formation of the A2029-BCG IC~1101}\label{Sec4.3}

Massive galaxies are thought to build up hierarchically when smaller
systems merge to form larger ones (\citealt{1977egsp.conf..401T};
\citealt{1978MNRAS.183..341W}). The high luminosities of BCGs, and
their distinct physical properties and location near the centre of
rich galaxy clusters suggest that they have experienced a higher
number of mergers and accretion events than other luminous galaxies
(e.g., \citealt{1978ApJ...224..320H}; \citealt{2007MNRAS.379..867V};
\citealt{2007MNRAS.375....2D}; \citealt{2007ApJ...665L...9R};
\citealt{2015MNRAS.449.3347O}). A2029, in which IC~1101 resides, is
the second richest cluster in the \citet[his
Table~3]{1978ApJ...226...55D} sample of 13 very rich Abell (1958)
clusters. \citet[his Fig.~1]{1978ApJ...226...55D} showed that A2029 and A154 are
the two clusters in his sample that have an excess number density
(over $R \la 4\arcmin$) with respect to the best King model fit to the
cluster spatial distribution of galaxies. \citet{1978ApJ...226...55D, 1978ApJ...223..765D,
  1979ApJ...231..659D} also found a deficit of
bright galaxies in the A2029 cluster, which led him to conclude that
most of the massive galaxies in A2029 had been captured by and had
merged with IC~1101, perhaps explaining the presence of the
intermediate-scale and outer exponential halo components of IC~1101
(Fig.~\ref{Fig5}). \citet[see also \citealt{1990dig..book..394T}]
{1985ApJ...289...18M}, however, argued that such cannibalism is less
efficient due to the clusters' high velocity dispersion (but see
\citealt{1988ApJ...325...49L}).

The findings by \citet{1978ApJ...226...55D, 1978ApJ...223..765D,
  1979ApJ...231..659D} may suggest that the extremely large core of
IC~1101 is somewhat special and that the galaxy may have undergone a
large number of dry major mergers involving binary SMBH
scouring. Nonetheless, $\mathcal{N} \ga 76$ seems unrealistically
large. In Section~\ref{Sec4.2}, we derived $\mathcal{N}$ assuming that
(i) the extrapolations of the $M_{\rm BH}-\sigma$, $M_{\rm BH}-L$ and
$M_{\rm BH}-M_{*}$ relations follow the sequence defined by the most
massive BCGs and (ii) the cumulative scouring action of SMBH binaries
is the sole mechanism that generated the large core in the galaxy. Not
to be confused with galaxies whose $M_{\rm BH}/M_{\rm sph}$ ratio was
over-estimated because their disk light was over-estimated, and thus
their spheroid light was underestimated (see
\citealt{2016MNRAS.457..320S}), recent studies have revealed that the
BHs in some BCGs are overmassive ($M_{\rm BH} \ga 10^{10} M_{\sun}$)
compared to those predicted from $M_{\rm BH}-\sigma$ and
$M_{\rm BH}-L$ relations (e.g., \citealt{2011Natur.480..215M};
\citealt{2016Natur.532..340T}).  Also, the extrapolation of the
relation between the black hole mass and the break radius
($M_{\rm BH}-R_{\rm b}$) for core-S\'ersic galaxies (e.g.,
\citealt[their eqs.\ 15, 16 and 17]{2014MNRAS.444.2700D};
\citealt[their eq.\ 13]{2013AJ....146..160R}; \citealt[their Fig.\
4]{2016Natur.532..340T}) is such that IC~1101 would host a SMBH with
$M_{\rm BH} \sim (4-10)\times10^{10} M_{\sun}$ (see also
\citealt{2015MNRAS.451.1177L}) $\approx$ (1.7 --
3.2$)\times\sigma_{s}$ (rms scatter) larger than those estimated by
the $M_{\rm BH}-\sigma$, $M_{\rm BH}-L$ and $M_{\rm BH}-M_{*}$
relations, thereby reducing the required merger rate for the galaxy to
10 -- 20 major mergers, if the core depletion is only due to binary
SMBHs. This possible existence of an overmassive black hole in IC~1101
supports the recent picture that the $M_{\rm BH}-\sigma$ and
$M_{\rm BH}-L$ relations for elliptical galaxies and bulges may
underestimate the SMBH masses of BCGs (e.g.,
\citealt{2011Natur.480..215M}; \citealt{2012MNRAS.424..224H};
\citealt{2016Natur.532..340T}).

Additional mechanisms, aside from binary SMBH scouring, may enhance a
pre-existing depleted core and produce the large
$M_{\rm def}/M_{\rm BH}$ ratio in IC~1101. For example,
gravitationally ``recoiled'' SMBHs, and the ensuing repeated
oscillation about the centre of a ``dry'' major merger remnant is
invoked to explain $M_{\rm def}/M_{\rm BH} \sim 5$ rather than 0.5
(e.g., \citealt{1989ComAp..14..165R}; \citealt{2004ApJ...613L..37B};
\citealt{2004ApJ...607L...9M};
\citealt{2008ApJ...678..780G}). Therefore, if this process has
occurred, then $\mathcal{N} \la 10$ instead of 10 $-$ 20 for
IC~1101. In Fig.~\ref{FigRes} (left), the residual structure reveals
that the core of IC~1101 is slightly offset with respect to the outer
isophotes, in agreement with the recoiled black hole oscillation
scenario (see also \citealt[their Fig.~3]{2012ApJ...756..159P}).

We postulate that the intermediate-scale component and the outer halo
of IC~1101 are in part consequences of galaxy merging and/or accretion
of stars that were tidal stripped from less massive galaxies (e.g.,
\citealt{1972AJ.....77..288G}). Furthermore, star formation in a
cooling flow can add young stars to the halo of IC~1101.
\citet{2004ApJ...616..178C} found excess X-ray emission and high
cooling flow rates in the A2029 cluster, concluding that the gas in
the cluster has recently started cooling and forming stars (see also
\citealt{2012MNRAS.422.3503W} and \citealt{2013ApJ...773..114P}).  In
addition, we find that the ejection of inner stars from the core of
IC~1101 due to coalescing SMBHs and other mechanisms have created an
extremely large stellar light deficit of log~($L_{\rm def}/L_{\sun}$)
$\sim$ 11.04, i.e., a $V$-band absolute magnitude of
$M_{V,\rm def} \sim -22.8$ mag $\approx$ 1/3 of the galaxy's spheroid
luminosity prior to the scouring (Section~\ref{Sec4.2}).  This high
stellar light deficit for IC~1101 implies that ejected stars, which
accumulate at large radii outside the core or escape from the galaxy
at high velocities (\citealt{1988Natur.331..687H} ), can impact the
light profile outside the core region. This partly explains the
presence of the intermediate-scale component with $V$-band absolute
magnitude of $M_{V} \sim -23.4$ mag.

Overall, the large depleted core, ellipticity, orientation, the
boxiness/disciness parameter ($B_{4}$), the relatively blue colour of
the BCG at large radii (Figs.~\ref{Fig4}, \ref{Fig5} and \ref{Fig7}),
and the velocity dispersion profile (\citealt{1979ApJ...231..659D})
favor the build-up of IC~1101 via a reasonably large number of dry
major mergers ($\mathcal{N} \la 10$) involving SMBHs (with the final
merged SMBH mass $M_{\rm BH} \ga 10^{10} M_{\sun}$) and a number of
accretion events. In this scenario, the central SMBH tidally disrupts
any survived stellar nuclei of accreted/merged satellites, which
otherwise would refill the depleted core
(\citealt{1997AJ....114.1771F}; \citealt{2007MNRAS.374.1227B};
\citealt{2010ApJ...714L.313B}). Our results suggest that the extremely
large partially-depleted core and $M_{\rm def}/M_{\rm BH}$ of IC~1101
maybe partly due to the actions of oscillatory core-passages by a
gravitational wave-kicked SMBH.

\section{Conclusions}\label{ConV} 

We have provided a detailed discussion of a newly discovered and
extremely large ($R_{\rm b} \sim$ $2\farcs77 \approx$ 4.2 $\pm$ 0.1
kpc) partially-depleted core in the A2029 BCG IC~1101. Extracting the
1D surface brightness profile of IC~1101 from high-resolution {\it
  HST} WFPC2 imaging, we perform a careful four-component
decomposition into a small elongated Gaussian nucleus, a core-S\'ersic
spheroid, a S\'ersic intermediate-scale component and an exponential
stellar halo. We also perform a 2D spheroid +
intermediate-scale component + stellar halo + nucleus decomposition
of IC~1101's {\it HST} WFPC2 image using {\sc imfit}. The main
findings from this work are as follows.\\

(1) The 1D spheroid + intermediate-scale component +
stellar halo + nucleus model yields an excellent fit to the BCG's
light profile. The rms residual scatter is below 0.02 mag
arcsec$^{-2}$. We found good agreement between the 1D and 2D
decompositions. In contrast, a 1D Gaussian nuclear
component + S\'ersic spheroid + S\'ersic intermediate-scale component +
exponential stellar halo model does not fit the light profile well,
instead this fit creates a residual structure revealing the
partially-depleted core of the galaxy.

(2) IC~1101 has the largest core size (i.e., measured by the
core-S\'ersic break radius $R_{\rm b} \approx$ 4.2 $\pm$ 0.1 kpc) to
date.  This break radius is an order of magnitude larger than those
typically measured for core-S\'ersic galaxies (i.e., $R_{\rm b}\sim$
20 pc $-$ 500 pc, e.g., \citealt{2006ApJS..164..334F};
\citealt{2011MNRAS.415.2158R}; \citealt{2013AJ....146..160R};
\citealt{2014MNRAS.444.2700D}). For comparison,
\citet{2012ApJ...756..159P} modeled the semi-major axis profile of
A2261-BCG and found the hitherto largest core ($r_{\gamma=0.5} \sim$
3.2 $\pm$ 0.1 kpc), this large core was recently confirmed by
\citet{2016ApJ...829...81B} who measured $R_{\rm b} \sim 3.6$ kpc.

(3) This depleted core in IC~1101 follows the extrapolation of the
$R_{\rm b}$$-$ $\mu_{\rm
  b}$ relation for core-S\'ersic galaxies (e.g.,
\citealt{1997AJ....114.1771F};
\citealt{2014MNRAS.444.2700D}). However, the spheroid contains $\sim
$ 25\% of the total galaxy light and has a
$V$-band absolute magnitude ($M_{V}$) of
$-23.8$ mag, which is faint for the large $R_{\rm
  b}$. As such, the observed depleted core of IC~1101 is 1.02 dex
$\approx$
3.4$\sigma_{s}$ (rms scatter) larger than that estimated from the
$R_{\rm b} -M_{V}$ relation (\citealt{2014MNRAS.444.2700D}).

(4) We measured a stellar mass deficit at the centre of IC~1101
$\approx$ $M_{\rm def}$ $\sim$ $4.9 \times 10^{11}
M_{\sun}$, (a luminosity deficit $L_{\rm def}/L_{\sun} \sim
1.1\times10^{11} $ in the $F702W$-band
$\approx$ 28\% of the spheroid luminosity before the core
depletion). Estimating the black hole mass of the galaxy using the
spheroid's stellar mass ($M_{*} \sim 1.1 \times 10^{12}
M_{\sun}$), luminosity ($M_{V} \sim
-23.8$ mag) and velocity dispersion ($\sigma \sim 378$ km
s$^{-1}$) yields $M_{\rm def}/M_{\rm BH}$ ratios of
$\sim$ 38, 50 and 73
$-$ 101, respectively.  This figure translates to spheroid formation
via an unrealistically large ($\mathcal{N} \ga
76$) number of ``dry" major mergers. However, the extrapolation of the
relation between the black hole mass and the break radius ($M_{\rm
  BH}-R_{\rm
  b}$) for core-S\'ersic galaxies (\citealt{2013AJ....146..160R};
\citealt{2014MNRAS.444.2700D}; \citealt{2016Natur.532..340T}) suggests
IC~1101 hosts an overmassive BH ($M_{\rm BH} \sim (4 -10) \times
10^{10} M_{\sun}$) $\approx$ $(1.7 - 3.2)\times
\sigma_{s}$ (rms scatter) larger than those SMBH masses estimated by
the $M_{\rm BH}-\sigma$ and $M_{\rm
  BH}-L$ relations, thereby reducing the merger rate for the galaxy to
$\mathcal{N} \la
10$, in close agreement with observational and theoretical merger
rates of massive galaxies. An additional mechanism that can contribute
to the large core/mass deficit is oscillatory core passages by a
recoiled SMBH. It is important to investigate the reasons why BCGs
with extremely large depleted cores and stellar mass deficits similar
to IC~1101 are quite rare, especially within a distance of $\sim$100
Mpc.

\section{ACKNOWLEDGMENTS}

BTD thanks the referee for their careful reading of the paper and
useful suggestions.  BTD acknowledges support from a Spanish
postdoctoral fellowship ``Ayudas para la atracci\'on del talento
investigador. Modalidad 2: j\'ovenes investigadores, financiadas por
la Comunidad de Madrid'' under grant number 2016-T2/TIC-2039.  BTD \&
JHK acknowledge financial support from the Spanish Ministry of Economy
and Competitiveness (MINECO) under grant number AYA2013-41243-P. JHK
acknowledges financial support from the European Union’s Horizon 2020
research and innovation programme under Marie Skłodowska-Curie grant
agreement No 721463 to the SUNDIAL ITN network, and from the Spanish
Ministry of Economy and Competitiveness (MINECO) under grant number
AYA2016-76219-P. AWG was supported under the Australian Research
Council's funding scheme (DP17012923).  JHK thanks the Astrophysics
Research Institute of Liverpool John Moores University for their
hospitality, and the Spanish Ministry of Education, Culture and Sports
for financial support of his visit there, through grant number
PR2015-00512. This research made use of APLpy, an open-source plotting
package for Python (\citealt{2012ascl.soft08017R})

\bibliographystyle{mnras}

\begin{thebibliography}{}
\bibitem[\protect\citeauthoryear{Abell}{1958}]{1958ApJS....3..211A} Abell G.~O., 1958, ApJS, 3, 211
\bibitem[\protect\citeauthoryear{Antonini \& Merritt}{2012}]{2012ApJ...745...83A} Antonini F., Merritt D., 2012, ApJ, 745, 83 
\bibitem[\protect\citeauthoryear{Arca-Sedda et al.}{2015}]{2015ApJ...806..220A} Arca-Sedda M., Capuzzo-Dolcetta R., Antonini F., Seth A., 2015, ApJ, 806, 220 
\bibitem[\protect\citeauthoryear{Arca-Sedda, Capuzzo-Dolcetta, \& Spera}{2016}]{2016MNRAS.456.2457A} Arca-Sedda M., Capuzzo-Dolcetta R., Spera M., 2016, MNRAS, 456, 2457 
\bibitem[\protect\citeauthoryear{Begelman, Blandford, \& Rees}{1980}]{1980Natur.287..307B} Begelman M.~C., Blandford R.~D., Rees M.~J., 1980, Natur, 287, 307 
\bibitem[\protect\citeauthoryear{Bekenstein}{1973}]{1973ApJ...183..657B} Bekenstein J.~D., 1973, ApJ, 183, 657 
\bibitem[\protect\citeauthoryear{Bekki \& Graham}{2010}]{2010ApJ...714L.313B} Bekki K., Graham A.~W., 2010, ApJ, 714, L313
\bibitem[\protect\citeauthoryear{Bell et al.}{2006}]{2006ApJ...640..241B} Bell E.~F., et al., 2006, ApJ, 640, 241 
\bibitem[\protect\citeauthoryear{Bender et al.}{2015}]{2015ApJ...807...56B} Bender R., Kormendy J., Cornell M.~E., Fisher D.~B., 2015, ApJ, 807, 56 
\bibitem[\protect\citeauthoryear{Bertin \& Arnouts}{1996}]{1996A&AS..117..393B} Bertin E., Arnouts S., 1996, A\&AS, 117, 393 
\bibitem[\protect\citeauthoryear{Binney \& Mamon}{1982}]{1982MNRAS.200..361B} Binney J., Mamon G.~A., 1982, MNRAS, 200, 361 
\bibitem[\protect\citeauthoryear{Blanton et al.}{2011}]{2011AJ....142...31B} Blanton M.~R., Kazin E., Muna D., Weaver B.~A., Price-Whelan A., 2011, AJ, 142, 31
\bibitem[\protect\citeauthoryear{Bluck et al.}{2012}]{2012ApJ...747...34B} Bluck A.~F.~L., Conselice C.~J., Buitrago F., Gr{\"u}tzbauch R., Hoyos C., Mortlock A., Bauer A.~E., 2012, ApJ, 747, 34 
\bibitem[\protect\citeauthoryear{Bonfini}{2014}]{2014PASP..126..935B} Bonfini P., 2014, PASP, 126, 935 
\bibitem[\protect\citeauthoryear{Bonfini, Dullo, \& Graham}{2015}]{2015ApJ...807..136B} Bonfini P., Dullo B.~T., Graham A.~W., 2015, ApJ, 807, 136 
\bibitem[\protect\citeauthoryear{Bonfini \& Graham}{2016}]{2016ApJ...829...81B} Bonfini P., Graham A.~W., 2016, ApJ, 829, 81
\bibitem[\protect\citeauthoryear{Boylan-Kolchin \& Ma}{2007}]{2007MNRAS.374.1227B} Boylan-Kolchin M., Ma C.-P., 2007, MNRAS, 374, 1227 
\bibitem[\protect\citeauthoryear{Boylan-Kolchin, Ma, \& Quataert}{2004}]{2004ApJ...613L..37B} Boylan-Kolchin M., Ma C.-P., Quataert E., 2004, ApJ, 613, L37 
\bibitem[\protect\citeauthoryear{Buote \& Tsai}{1996}]{1996ApJ...458...27B} Buote D.~A., Tsai J.~C., 1996, ApJ, 458, 27 
\bibitem[\protect\citeauthoryear{Byun et al.}{1996}]{1996AJ....111.1889B} Byun Y.-I., et al., 1996, AJ, 111, 1889 
\bibitem[\protect\citeauthoryear{Carter}{1977}]{1977MNRAS.178..137C} Carter D., 1977, MNRAS, 178, 137 
\bibitem[\protect\citeauthoryear{Carter, Bridges, \& Hau}{1999}]{1999MNRAS.307..131C} Carter D., Bridges T.~J., Hau G.~K.~T., 1999, MNRAS, 307, 131 
\bibitem[\protect\citeauthoryear{Carter et al.}{1985}]{1985MNRAS.212..471C} Carter D., Inglis I., Ellis R.~S., Efstathiou G., Godwin J.~G., 1985, MNRAS, 212, 471
\bibitem[\protect\citeauthoryear{Carollo et al.}{1997}]{1997ApJ...481..710C} Carollo C.~M., Franx M., Illingworth G.~D., Forbes D.~A., 1997, ApJ, 481, 710
\bibitem[\protect\citeauthoryear{Casteels et al.}{2014}]{2014MNRAS.445.1157C} Casteels K.~R.~V., et al., 2014, MNRAS, 445, 1157 
\bibitem[\protect\citeauthoryear{Ciambur}{2016}]{2016PASA...33...62C} Ciambur B.~C., 2016, PASA, 33, e062 
\bibitem[\protect\citeauthoryear{Clarke, Blanton, \& Sarazin}{2004}]{2004ApJ...616..178C} Clarke T.~E., Blanton E.~L., Sarazin C.~L., 2004, ApJ, 616, 178
\bibitem[\protect\citeauthoryear{Condon et al.}{1998}]{1998AJ....115.1693C} Condon J.~J., Cotton W.~D., Greisen E.~W., Yin Q.~F., Perley R.~A., Taylor G.~B., Broderick J.~J., 1998, AJ, 115, 1693 
\bibitem[\protect\citeauthoryear{Conselice}{2007}]{2007IAUS..235..381C} Conselice C.~J., 2007, IAUS, 235, 381 
\bibitem[\protect\citeauthoryear{Crane et al.}{1993}]{1993AJ....106.1371C} Crane P., et al., 1993, AJ, 106, 1371 
\bibitem[\protect\citeauthoryear{Debattista et al.}{2006}]{2006ApJ...651L..97D} Debattista V.~P., Ferreras I., Pasquali A., Seth A., De Rijcke S., Morelli L., 2006, ApJ, 651, L97 
\bibitem[\protect\citeauthoryear{De Lucia \& Blaizot}{2007}]{2007MNRAS.375....2D} De Lucia G., Blaizot J., 2007, MNRAS, 375, 2
\bibitem[\protect\citeauthoryear{de Vaucouleurs}{1948}]{1948AnAp...11..247D} de Vaucouleurs G., 1948, AnAp, 11, 247 
\bibitem[\protect\citeauthoryear{de Vaucouleurs et al.}{1991}]{1991rc3..book.....D} de Vaucouleurs G., de Vaucouleurs A., Corwin H.~G., Jr., Buta R.~J., Paturel G., Fouqu{\'e} P., 1991, rc3..book, I,  
\bibitem[\protect\citeauthoryear{Donzelli, Muriel, \& Madrid}{2011}]{2011ApJS..195...15D} Donzelli C.~J., Muriel H., Madrid J.~P., 2011, ApJS, 195, 15 
\bibitem[\protect\citeauthoryear{Dullo \& Graham}{2012}]{2012ApJ...755..163D} Dullo B.~T., Graham A.~W., 2012, ApJ, 755, 163 
\bibitem[\protect\citeauthoryear{Dullo \& Graham}{2013}]{2013ApJ...768...36D} Dullo B.~T., Graham A.~W., 2013, ApJ, 768, 36 
\bibitem[\protect\citeauthoryear{Dullo \& Graham}{2014}]{2014MNRAS.444.2700D} Dullo B.~T., Graham A.~W., 2014, MNRAS, 444, 2700 
\bibitem[\protect\citeauthoryear{Dullo \& Graham}{2015}]{2015ApJ...798...55D} Dullo B.~T., Graham A.~W., 2015, ApJ, 798, 55 
\bibitem[\protect\citeauthoryear{Dullo, Mart{\'{\i}}nez-Lombilla, \&
    Knapen}{2016}]{2016MNRAS.462.3800D} Dullo B.~T.,
  Mart{\'{\i}}nez-Lombilla C., Knapen J.~H., 2016, MNRAS, 462, 3800 
\bibitem[\protect\citeauthoryear{Dressler}{1978a}]{1978ApJ...226...55D} Dressler A., 1978, ApJ, 226, 55a
\bibitem[\protect\citeauthoryear{Dressler}{1978b}]{1978ApJ...223..765D} Dressler A., 1978, ApJ, 223, 765b 
\bibitem[\protect\citeauthoryear{Dressler}{1979}]{1979ApJ...231..659D} Dressler A., 1979, ApJ, 231, 659 
\bibitem[\protect\citeauthoryear{Dressler}{1981}]{1981ApJ...243...26D} Dressler A., 1981, ApJ, 243, 26 
\bibitem[\protect\citeauthoryear{Dressler}{1989}]{1989IAUS..134..217D} Dressler A., 1989, IAUS, 134, 217 
\bibitem[\protect\citeauthoryear{Ebisuzaki, Makino, \& Okumura}{1991}]{1991Natur.354..212E} Ebisuzaki T., Makino J., Okumura S.~K., 1991, Natur, 354, 212 
\bibitem[\protect\citeauthoryear{Erwin}{2015}]{2015ApJ...799..226E} Erwin P., 2015, ApJ, 799, 226 
\bibitem[\protect\citeauthoryear{Faber et al.}{1997}]{1997AJ....114.1771F} Faber S.~M., et al., 1997, AJ, 114, 1771 
\bibitem[\protect\citeauthoryear{Ferrarese et al.}{2006}]{2006ApJS..164..334F} Ferrarese L., et al., 2006, ApJS, 164, 334 
\bibitem[\protect\citeauthoryear{Ferrarese \& Merritt}{2000}]{2000ApJ...539L...9F} Ferrarese L., Merritt D., 2000, ApJ, 539, L9 
\bibitem[\protect\citeauthoryear{Ferrarese et al.}{1994}]{1994AJ....108.1598F} Ferrarese L., van den Bosch F.~C., Ford H.~C., Jaffe W., O'Connell R.~W., 1994, AJ, 108, 1598 
\bibitem[\protect\citeauthoryear{Fisher, Illingworth, \& Franx}{1995}]{1995ApJ...438..539F} Fisher D., Illingworth G., Franx M., 1995, ApJ, 438, 539 
\bibitem[\protect\citeauthoryear{Fitchett}{1983}]{1983MNRAS.203.1049F} Fitchett M.~J., 1983, MNRAS, 203, 1049 
\bibitem[\protect\citeauthoryear{Forbes, Franx, \& Illingworth}{1995}]{1995AJ....109.1988F} Forbes D.~A., Franx M., Illingworth G.~D., 1995, AJ, 109, 1988 
\bibitem[\protect\citeauthoryear{Fukugita, Shimasaku, \& Ichikawa}{1995}]{1995PASP..107..945F} Fukugita M., Shimasaku K., Ichikawa T., 1995, PASP, 107, 945 
\bibitem[\protect\citeauthoryear{Gallagher \& Ostriker}{1972}]{1972AJ.....77..288G} Gallagher J.~S., III, Ostriker J.~P., 1972, AJ, 77, 288 
\bibitem[\protect\citeauthoryear{Gebhardt et al.}{2000}]{2000ApJ...539L..13G} Gebhardt K., et al., 2000, ApJ, 539, L13 
\bibitem[\protect\citeauthoryear{Gebhardt et al.}{1996}]{1996AJ....112..105G} Gebhardt K., et al., 1996, AJ, 112, 105 
\bibitem[\protect\citeauthoryear{Gebhardt et al.}{2003}]{2003ApJ...583...92G} Gebhardt K., et al., 2003, ApJ, 583, 92 
\bibitem[\protect\citeauthoryear{Goerdt et al.}{2010}]{2010ApJ...725.1707G} Goerdt T., Moore B., Read J.~I., Stadel J., 2010, ApJ, 725, 1707 
\bibitem[\protect\citeauthoryear{Gonzalez, Zabludoff, \& Zaritsky}{2003}]{2003Ap&SS.285...67G} Gonzalez A.~H., Zabludoff A.~I., Zaritsky D., 2003, Ap\&SS, 285, 67 
\bibitem[\protect\citeauthoryear{Gonzalez, Zabludoff, \& Zaritsky}{2005}]{2005ApJ...618..195G} Gonzalez A.~H., Zabludoff A.~I., Zaritsky D., 2005, ApJ, 618, 195 
%\bibitem[\protect\citeauthoryear{Graham}{2001}]{2001AJ....121..820G} Graham A.~W., 2001, AJ, 121, 820 
\bibitem[\protect\citeauthoryear{Graham}{2004}]{2004ApJ...613L..33G} Graham A.~W., 2004, ApJ, 613, L33 
%\bibitem[\protect\citeauthoryear{Graham, Ciambur, \& Savorgnan}{2016}]{2016ApJ...831..132G} Graham A.~W., Ciambur B.~C., Savorgnan G.~A.~D., 2016, ApJ, 831, 132 
\bibitem[\protect\citeauthoryear{Graham \& Driver}{2005}]{2005PASA...22..118G} Graham A.~W., Driver S.~P., 2005, PASA, 22, 118 
\bibitem[\protect\citeauthoryear{Graham et al.}{2003}]{2003AJ....125.2951G} Graham A.~W., Erwin P., Trujillo I., Asensio Ramos A., 2003, AJ, 125, 2951 
\bibitem[\protect\citeauthoryear{Graham \& Scott}{2013}]{2013ApJ...764..151G} Graham A.~W., Scott N., 2013, ApJ, 764, 151 
\bibitem[\protect\citeauthoryear{Graham \& Spitler}{2009}]{2009MNRAS.397.2148G} Graham A.~W., Spitler L.~R., 2009, MNRAS, 397, 2148 
\bibitem[\protect\citeauthoryear{Grillmair et al.}{1994}]{1994AJ....108..102G} Grillmair C.~J., Faber S.~M., Lauer T.~R., Baum W.~A., Lynds R.~C., O'Neil E.~J., Jr., Shaya E.~J., 1994, AJ, 108, 102
\bibitem[\protect\citeauthoryear{Gualandris \& Merritt}{2008}]{2008ApJ...678..780G} Gualandris A., Merritt D., 2008, ApJ, 678, 780-797 
\bibitem[\protect\citeauthoryear{Haehnelt \& Kauffmann}{2002}]{2002MNRAS.336L..61H} Haehnelt M.~G., Kauffmann G., 2002, MNRAS, 336, L61 
\bibitem[\protect\citeauthoryear{Hausman \& Ostriker}{1978}]{1978ApJ...224..320H} Hausman M.~A., Ostriker J.~P., 1978, ApJ, 224, 320 
\bibitem[\protect\citeauthoryear{Hills}{1988}]{1988Natur.331..687H} Hills J.~G., 1988, Natur, 331, 687 
\bibitem[\protect\citeauthoryear{Hlavacek-Larrondo et al.}{2012}]{2012MNRAS.424..224H} Hlavacek-Larrondo J., Fabian A.~C., Edge A.~C., Hogan M.~T., 2012, MNRAS, 424, 224 
\bibitem[\protect\citeauthoryear{Hopkins \& Hernquist}{2010}]{2010MNRAS.407..447H} Hopkins P.~F., Hernquist L., 2010, MNRAS, 407, 447  
\bibitem[\protect\citeauthoryear{Hyde et al.}{2008}]{2008MNRAS.391.1559H} Hyde J.~B., Bernardi M., Sheth R.~K., Nichol R.~C., 2008, MNRAS, 391, 1559 
\bibitem[\protect\citeauthoryear{Jaffe et al.}{1994}]{1994AJ....108.1567J} Jaffe W., Ford H.~C., O'Connell R.~W., van den Bosch F.~C., Ferrarese L., 1994, AJ, 108, 1567 
\bibitem[\protect\citeauthoryear{Jedrzejewski}{1987}]{1987MNRAS.226..747J} Jedrzejewski R.~I., 1987, MNRAS, 226, 747 
\bibitem[\protect\citeauthoryear{Kelson et al.}{2002}]{2002ApJ...576..720K} Kelson D.~D., Zabludoff A.~I., Williams K.~A., Trager S.~C., Mulchaey J.~S., Bolte M., 2002, ApJ, 576, 720 
\bibitem[\protect\citeauthoryear{Khosroshahi, Ponman, \& Jones}{2006}]{2006MNRAS.372L..68K} Khosroshahi H.~G., Ponman T.~J., Jones L.~R., 2006, MNRAS, 372, L68
\bibitem[\protect\citeauthoryear{King}{1978}]{1978ApJ...222....1K} King I.~R., 1978, ApJ, 222, 1 
\bibitem[\protect\citeauthoryear{King \& Minkowski}{1966}]{1966ApJ...143.1002K} King I.~R., Minkowski R., 1966, ApJ, 143, 1002 
\bibitem[\protect\citeauthoryear{Kormendy et al.}{1994}]{1994ESOC...49..147K} Kormendy J., Dressler A., Byun Y.~I., Faber S.~M., Grillmair C., Lauer T.~R., Richstone D., Tremaine S., 1994, ESOC, 49, 147 
\bibitem[\protect\citeauthoryear{Kormendy \& Bender}{2009}]{2009ApJ...691L.142K} Kormendy J., Bender R., 2009, ApJ, 691, L142 
\bibitem[\protect\citeauthoryear{Kormendy \& Ho}{2013}]{2013ARA&A..51..511K} Kormendy J., Ho L.~C., 2013, ARA\&A, 51, 511 
\bibitem[\protect\citeauthoryear{Kormendy \& Richstone}{1995}]{1995ARA&A..33..581K} Kormendy J., Richstone D., 1995, ARA\&A, 33, 581 
\bibitem[\protect\citeauthoryear{Krist}{1995}]{1995ASPC...77..349K} Krist J., 1995, ASPC, 77, 349 
\bibitem[\protect\citeauthoryear{Kulkarni \& Loeb}{2012}]{2012MNRAS.422.1306K} Kulkarni G., Loeb A., 2012, MNRAS, 422, 1306 
\bibitem[\protect\citeauthoryear{Laine et al.}{2003}]{2003AJ....125..478L} Laine S., van der Marel R.~P., Lauer T.~R., Postman M., O'Dea C.~P., Owen F.~N., 2003, AJ, 125, 478 
\bibitem[\protect\citeauthoryear{Laporte \& White}{2015}]{2015MNRAS.451.1177L} Laporte C.~F.~P., White S.~D.~M., 2015, MNRAS, 451, 1177 
\bibitem[\protect\citeauthoryear{Lauer}{1988}]{1988ApJ...325...49L} Lauer T.~R., 1988, ApJ, 325, 49 
\bibitem[\protect\citeauthoryear{Lauer et al.}{1995}]{1995AJ....110.2622L} Lauer T.~R., et al., 1995, AJ, 110, 2622 
\bibitem[\protect\citeauthoryear{Lauer et al.}{1996}]{1996ApJ...471L..79L} Lauer T.~R., et al., 1996, ApJ, 471, L79
\bibitem[\protect\citeauthoryear{Lauer et al.}{2005}]{2005AJ....129.2138L} Lauer T.~R., et al., 2005, AJ, 129, 2138 
\bibitem[\protect\citeauthoryear{Lauer et al.}{2007}]{2007ApJ...662..808L} Lauer T.~R., et al., 2007, ApJ, 662, 808 
\bibitem[\protect\citeauthoryear{L{\'o}pez-Cruz et al.}{2014}]{2014ApJ...795L..31L} L{\'o}pez-Cruz O., A{\~n}orve C., Birkinshaw M., Worrall D.~M., Ibarra-Medel H.~J., Barkhouse W.~A., Torres-Papaqui J.~P., Motta V., 2014, ApJ, 795, L31 
\bibitem[\protect\citeauthoryear{Lugger}{1984}]{1984ApJ...286..106L} Lugger P.~M., 1984, ApJ, 286, 106 
\bibitem[\protect\citeauthoryear{Madrid \& Donzelli}{2016}]{2016ApJ...819...50M} Madrid J.~P., Donzelli C.~J., 2016, ApJ, 819, 50 
\bibitem[\protect\citeauthoryear{Magorrian et al.}{1998}]{1998AJ....115.2285M} Magorrian J., et al., 1998, AJ, 115, 2285
\bibitem[\protect\citeauthoryear{Matthews, Morgan, \& Schmidt}{1964}]{1964ApJ...140...35M} Matthews T.~A., Morgan W.~W., Schmidt M., 1964, ApJ, 140, 35 
\bibitem[\protect\citeauthoryear{Mazzalay et al.}{2016}]{2016MNRAS.462.2847M} Mazzalay X., Thomas J., Saglia R.~P., Wegner G.~A., Bender R., Erwin P., Fabricius M.~H., Rusli S.~P., 2016, MNRAS, 462, 2847 
\bibitem[\protect\citeauthoryear{McConnell \& Ma}{2013}]{2013ApJ...764..184M} McConnell N.~J., Ma C.-P., 2013, ApJ, 764, 184 
\bibitem[\protect\citeauthoryear{McConnell et al.}{2011}]{2011Natur.480..215M} McConnell N.~J., Ma C.-P., Gebhardt K., Wright S.~A., Murphy J.~D., Lauer T.~R., Graham J.~R., Richstone D.~O., 2011, Natur, 480, 215 
\bibitem[\protect\citeauthoryear{Merritt}{1985}]{1985ApJ...289...18M} Merritt D., 1985, ApJ, 289, 18 
\bibitem[\protect\citeauthoryear{Merritt}{2006}]{2006ApJ...648..976M} Merritt D., 2006, ApJ, 648, 976 
\bibitem[\protect\citeauthoryear{Merritt et al.}{2004}]{2004ApJ...607L...9M} Merritt D., Milosavljevi{\'c} M., Favata M., Hughes S.~A., Holz D.~E., 2004, ApJ, 607, L9 
\bibitem[\protect\citeauthoryear{Milosavljevi{\'c} \& Merritt}{2001}]{2001ApJ...563...34M} Milosavljevi{\'c} M., Merritt D., 2001, ApJ, 563, 34 
\bibitem[\protect\citeauthoryear{Oemler}{1976}]{1976ApJ...209..693O} Oemler A., Jr., 1976, ApJ, 209, 693 
\bibitem[\protect\citeauthoryear{Oliva-Altamirano et al.}{2015}]{2015MNRAS.449.3347O} Oliva-Altamirano P., Brough S., Jimmy T., Kim-Vy, Couch W.~J., McDermid R.~M., Lidman C., von der Linden A., Sharp R., 2015, MNRAS, 449, 3347 
\bibitem[\protect\citeauthoryear{Paterno-Mahler et al.}{2013}]{2013ApJ...773..114P} Paterno-Mahler R., Blanton E.~L., Randall S.~W., Clarke T.~E., 2013, ApJ, 773, 114 
\bibitem[\protect\citeauthoryear{Paturel et al.}{2003}]{2003A&A...412...45P} Paturel G., Petit C., Prugniel P., Theureau G., Rousseau J., Brouty M., Dubois P., Cambr{\'e}sy L., 2003, A\&A, 412, 45 
\bibitem[\protect\citeauthoryear{Pierini et al.}{2008}]{2008A&A...483..727P} Pierini D., Zibetti S., Braglia F., B{\"o}hringer H., Finoguenov A., Lynam P.~D., Zhang Y.-Y., 2008, A\&A, 483, 727 
\bibitem[\protect\citeauthoryear{Porter, Schneider, \& Hoessel}{1991}]{1991AJ....101.1561P} Porter A.~C., Schneider D.~P., Hoessel J.~G., 1991, AJ, 101, 1561 
\bibitem[\protect\citeauthoryear{Postman \& Lauer}{1995}]{1995ApJ...440...28P} Postman M., Lauer T.~R., 1995, ApJ, 440, 28 
\bibitem[\protect\citeauthoryear{Postman et al.}{2012}]{2012ApJ...756..159P} Postman M., et al., 2012, ApJ, 756, 159 
\bibitem[\protect\citeauthoryear{Quinlan \& Hernquist}{1997}]{1997NewA....2..533Q} Quinlan G.~D., Hernquist L., 1997, NewA, 2, 533
\bibitem[\protect\citeauthoryear{Ravindranath et al.}{2001}]{2001AJ....122..653R} Ravindranath S., Ho L.~C., Peng C.~Y., Filippenko A.~V., Sargent W.~L.~W., 2001, AJ, 122, 653 
\bibitem[\protect\citeauthoryear{Redmount \& Rees}{1989}]{1989ComAp..14..165R} Redmount I.~H., Rees M.~J., 1989, ComAp, 14, 165 
\bibitem[\protect\citeauthoryear{Rest et al.}{2001}]{2001AJ....121.2431R} Rest A., van den Bosch F.~C., Jaffe W., Tran H., Tsvetanov Z., Ford H.~C., Davies J., Schafer J., 2001, AJ, 121, 2431 
\bibitem[\protect\citeauthoryear{Richings, Uttley, \& K{\"o}rding}{2011}]{2011MNRAS.415.2158R} Richings A.~J., Uttley P., K{\"o}rding E., 2011, MNRAS, 415, 2158 
\bibitem[\protect\citeauthoryear{Rines, Finn, \& Vikhlinin}{2007}]{2007ApJ...665L...9R} Rines K., Finn R., Vikhlinin A., 2007, ApJ, 665, L9 
\bibitem[\protect\citeauthoryear{Robitaille \& Bressert}{2012}]{2012ascl.soft08017R} Robitaille T., Bressert E., 2012, ascl.soft, ascl:1208.017 
\bibitem[\protect\citeauthoryear{Rusli et al.}{2013}]{2013AJ....146..160R} Rusli S.~P., Erwin P., Saglia R.~P., Thomas J., Fabricius M., Bender R., Nowak N., 2013, AJ, 146, 160 
%\bibitem[\protect\citeauthoryear{Saglia et al.}{1993}]{1993MNRAS.264..961S} Saglia R.~P., Bertschinger E., Baggley G., Burstein D., Colless M., Davies R.~L., McMahan R.~K., Jr., Wegner G., 1993, MNRAS, 264, 961 
\bibitem[\protect\citeauthoryear{Saglia et al.}{2016}]{2016ApJ...818...47S} Saglia R.~P., et al., 2016, ApJ, 818, 47 
\bibitem[\protect\citeauthoryear{Savorgnan et al.}{2016}]{2016ApJ...817...21S} Savorgnan G.~A.~D., Graham A.~W., Marconi A., Sani E., 2016, ApJ, 817, 21 
\bibitem[\protect\citeauthoryear{Savorgnan \& Graham}{2016}]{2016MNRAS.457..320S} Savorgnan G.~A.~D., Graham A.~W., 2016, MNRAS, 457, 320
\bibitem[\protect\citeauthoryear{Seigar, Graham, \&
    Jerjen}{2007}]{2007MNRAS.378.1575S} Seigar M.~S., Graham A.~W.,
  Jerjen H., 2007, MNRAS, 378, 1575 
\bibitem[\protect\citeauthoryear{Sembach \& Tonry}{1996}]{1996AJ....112..797S} Sembach K.~R., Tonry J.~L., 1996, AJ, 112, 797 
\bibitem[\protect\citeauthoryear{S{\'e}rsic}{1963}]{1963BAAA....6...41S} S{\'e}rsic J.~L., 1963, BAAA, 6, 41 
\bibitem[\protect\citeauthoryear{S{\'e}rsic}{1968}]{1968adga.book.....S}
  S\'ersic J.~L., 1968, Atlas de Galaxias Australes (C\'ordoba: Observatorio
Astronomico, Universidad Nacional de C\'ordoba)
\bibitem[\protect\citeauthoryear{Schombert}{1986}]{1986ApJS...60..603S} Schombert J.~M., 1986, ApJS, 60, 603 
\bibitem[\protect\citeauthoryear{Schombert}{1987}]{1987ApJS...64..643S} Schombert J.~M., 1987, ApJS, 64, 643
\bibitem[\protect\citeauthoryear{Stewart et al.}{1984}]{1984ApJ...285....1S} Stewart G.~C., Fabian A.~C., Jones C., Forman W., 1984, ApJ, 285, 1 
\bibitem[\protect\citeauthoryear{Thomas et al.}{2016}]{2016Natur.532..340T} Thomas J., Ma C.-P., McConnell N.~J., Greene J.~E., Blakeslee J.~P., Janish R., 2016, Natur, 532, 340
\bibitem[\protect\citeauthoryear{Thomas et al.}{2014}]{2014ApJ...782...39T} Thomas J., Saglia R.~P., Bender R., Erwin P., Fabricius M., 2014, ApJ, 782, 39 
\bibitem[\protect\citeauthoryear{Tonry}{1987}]{1987IAUS..127...89T}
  Tonry J.~L., 1987, IAUS, 127, 89 
\bibitem[\protect\citeauthoryear{Toomre}{1977}]{1977egsp.conf..401T} Toomre A., 1977, in The Evolution of Galaxies and Stellar Populations, ed. B. M. Tinsley \& R. B. Larson (New Haven, CT: Yale Univ. Press), 401
\bibitem[\protect\citeauthoryear{Tremaine}{1990}]{1990dig..book..394T} Tremaine S., 1990,  in Dynamics and Interactions of Galaxies, ed. R. Wielen
(Berlin: Springer), 394
\bibitem[\protect\citeauthoryear{Trujillo et al.}{2001a}]{2001MNRAS.321..269T} Trujillo I., Aguerri J.~A.~L., Cepa J., Guti{\'e}rrez C.~M., 2001, MNRAS, 321, 269a
\bibitem[\protect\citeauthoryear{Trujillo et al.}{2001b}]{2001MNRAS.328..977T} Trujillo I., Aguerri J.~A.~L., Cepa J., Guti{\'e}rrez C.~M., 2001, MNRAS, 328, 977b 
\bibitem[\protect\citeauthoryear{Trujillo et al.}{2004}]{2004AJ....127.1917T} Trujillo I., Erwin P., Asensio Ramos A., Graham A.~W., 2004, AJ, 127, 1917 
\bibitem[\protect\citeauthoryear{Uson, Boughn, \& Kuhn}{1991}]{1991ApJ...369...46U} Uson J.~M., Boughn S.~P., Kuhn J.~R., 1991, ApJ, 369, 46 
\bibitem[\protect\citeauthoryear{van den Bosch et al.}{1994}]{1994AJ....108.1579V} van den Bosch F.~C., Ferrarese L., Jaffe W., Ford H.~C., O'Connell R.~W., 1994, AJ, 108, 1579 
\bibitem[\protect\citeauthoryear{Vazdekis et al.}{2010}]{2010MNRAS.404.1639V} Vazdekis A., S{\'a}nchez-Bl{\'a}zquez P., Falc{\'o}n-Barroso J., Cenarro A.~J., Beasley M.~A., Cardiel N., Gorgas J., Peletier R.~F., 2010, MNRAS, 404, 1639
\bibitem[\protect\citeauthoryear{von der Linden et al.}{2007}]{2007MNRAS.379..867V} von der Linden A., Best P.~N., Kauffmann G., White S.~D.~M., 2007, MNRAS, 379, 867
\bibitem[\protect\citeauthoryear{Walker et al.}{2012}]{2012MNRAS.422.3503W} Walker S.~A., Fabian A.~C., Sanders J.~S., George M.~R., Tawara Y., 2012, MNRAS, 422, 3503 
\bibitem[\protect\citeauthoryear{Worthey}{1994}]{1994ApJS...95..107W} Worthey G., 1994, ApJS, 95, 107 
\bibitem[\protect\citeauthoryear{White \& Rees}{1978}]{1978MNRAS.183..341W} White S.~D.~M., Rees M.~J., 1978, MNRAS, 183, 341 
\bibitem[\protect\citeauthoryear{Xu et al.}{2012}]{2012ApJ...747...85X} Xu C.~K., Zhao Y., Scoville N., Capak P., Drory N., Gao Y., 2012, ApJ, 747, 85 
\bibitem[\protect\citeauthoryear{Young \& Currie}{1994}]{1994MNRAS.268L..11Y} Young C.~K., Currie M.~J., 1994, MNRAS, 268, L11 
\bibitem[\protect\citeauthoryear{Zibetti et al.}{2005}]{2005MNRAS.358..949Z} Zibetti S., White S.~D.~M., Schneider D.~P., Brinkmann J., 2005, MNRAS, 358, 949


\end{thebibliography}

\label{lastpage}\end{document}